\documentclass[%
 reprint,
 amsmath,amssymb,
 aps,
]{revtex4-1}

\usepackage[export]{adjustbox}
\usepackage{mwe}
\usepackage{graphicx,float,wrapfig}
\usepackage{amsmath}
\usepackage[left,switch,mathlines]{lineno}
\modulolinenumbers[2]

\usepackage{xcolor}
\usepackage{ulem}
\usepackage{graphicx}
\usepackage{dcolumn}
\usepackage{bm}

\usepackage{url}

\newcommand{\scaleVal}{0.25}

\usepackage{times}

\begin{document}


\title{Solving the quantum many-body problem via correlations measured with a momentum microscope}

\author{S.~S.~Hodgman$^{1}$}
\email{\normalsize{sean.hodgman@anu.edu.au}}

\author{R.~I.~Khakimov$^{1}$, R.~J.~Lewis-Swan$^{2,3}$, A.~G.~Truscott$^{1}$, and K.~V.~Kheruntsyan$^{2}$}

\affiliation{\normalsize{$^{1}$Research School of Physics and Engineering, Australian National University, Canberra 0200, Australia}}

\affiliation{\normalsize{$^{2}$University of Queensland, School of Mathematics and Physics, Brisbane, Queensland 4072, Australia}}

\affiliation{\normalsize{$^{3}$JILA, NIST and Department of Physics, University of Colorado, 440 UCB, Boulder, Colorado 80309, USA}}

\date{\today}

\begin{abstract}
In quantum many-body theory, all physical observables are described in terms of correlation functions between particle creation/annihilation operators. Measurement of such correlation functions can therefore be regarded as an operational solution to the quantum many-body problem. Here we demonstrate this paradigm by measuring multi-particle momentum correlations up to third order between ultracold helium atoms in an $s$-wave scattering halo of colliding Bose-Einstein condensates, using a quantum many-body momentum microscope. Our measurements allow us to extract a key building block of all higher-order correlations in this system---the pairing field amplitude. In addition, we demonstrate a record violation of the classical Cauchy-Schwarz inequality for correlated atom pairs and triples. Measuring multi-particle momentum correlations could provide new insights into effects such as unconventional superconductivity and many-body localisation.
\end{abstract}

\maketitle


In quantum physics, fully understanding and characterising complex systems, comprising a large (often macroscopic) number of interacting particles, is an extremely challenging problem. Solutions within the standard framework of (first-quantised) quantum mechanics generally require the knowledge of the full quantum many-body wavefunction.  This necessitates an exponentially large amount of information to be encoded and simulated using the many-body Schr\"{o}dinger equation. 
In an equivalent (second-quantised) quantum field theory formulation, the fundamental understanding of quantum many-body systems comes through the description of all physical observables via correlation functions between particle creation and annihilation operators. Here, the exponential complexity of the quantum many-body problem is converted into the need to know all possible multi-particle correlation functions, starting from two-, three-, and increasing to arbitrary $N$-particle (or higher-order) correlations.

\begin{figure}[bp]
	\centering
	\includegraphics[width=0.44\textwidth, keepaspectratio=true]
{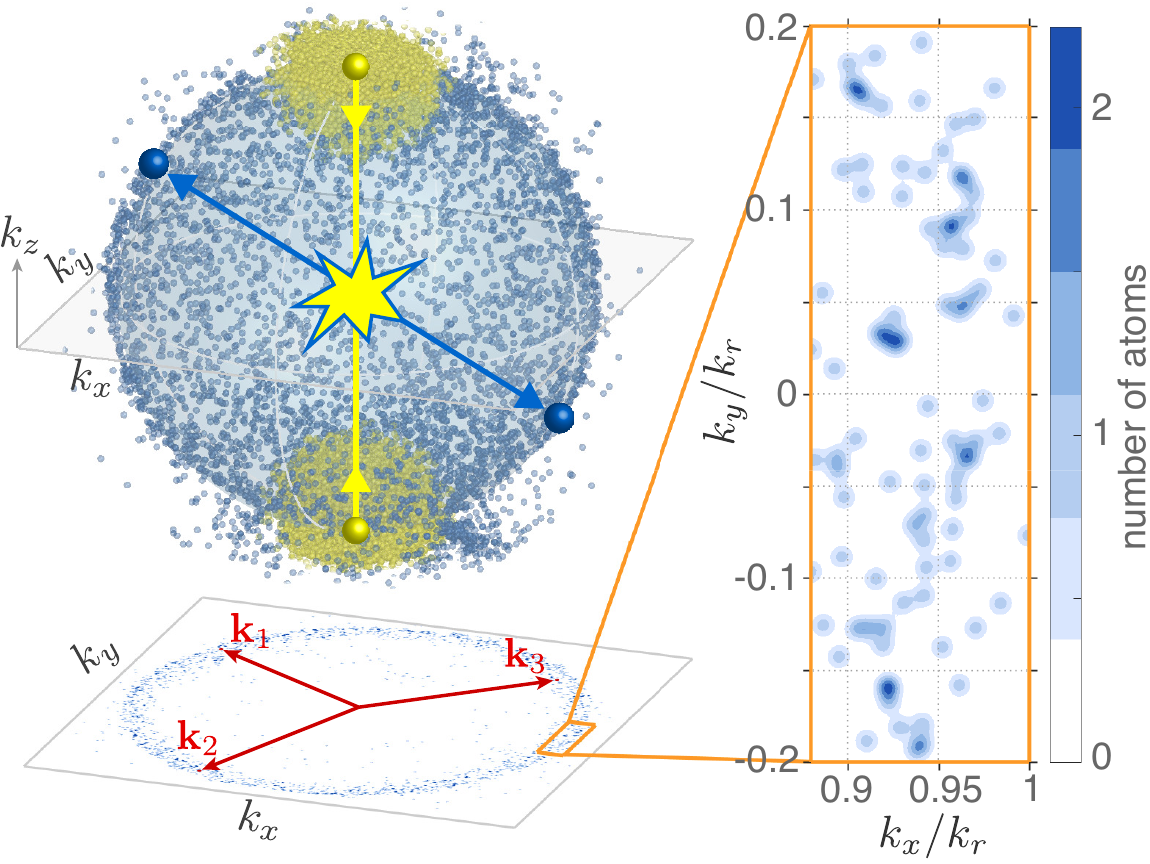}
	\caption{ {\bf Atomic momenta as measured by the quantum many-body momentum microscope.}  Individual momenta of detected atoms are reconstructed in 3D momentum space, with the main image showing the collision halo, with dense (yellow) patches on the north and south poles showing unscattered atoms from the pair of colliding condensates. The highlighted balls and arrows are an illustration of the underlying microscopic interactions---the binary $s$-wave collisions.
	The 2D histograms below show an equatorial slice through the experimental data, where the red arrows $\mathbf{k}_1,\mathbf{k}_2$, and $\mathbf{k}_3$ indicate three arbitrarily chosen momenta for which, e.g., three-atom correlations can be analysed via coincidence counts.
	Experimental data from 10 runs is shown, which approximates the density present in a single halo, given our detection efficiency of $\sim \!10$\%.  Individual atoms can be seen in the magnified inset, represented as 2D Gaussians with a width equal to the detector resolution. The size of the balls representing the individual atoms on the main 3D image are not to scale.  
	}
	\label{fig:schematic}
\end{figure} 

From an experimental viewpoint, an operational solution to the quantum many-body problem is therefore equivalent to measuring all multi-particle correlations. 
In certain cases, however, knowing only a specific set of (few-body or lower-order) correlations is sufficient to allow a solution of the many-body problem to be constructed. 
This was recently shown for phase correlations between two coupled one-dimensional (1D) Bose gases~\cite{Schweigler2016}.  
Apart from facilitating the description of physical observables, characterising multi-particle correlations is  important for introducing controlled approximations in many-body physics, such as the virial- and related cluster-expansion approaches that rely on truncation of the Bogolyubov-Born-Green-Kirkwood-Yvon hierarchy ~\cite{shavitt2009many,kira2011semiconductor}.
Momentum correlations up to 6th order~\cite{Dall2013_idealnbody} and phase correlations up to 8th  \cite{Schweigler2016} and 10th order \cite{Langen2015} have so far been measured in ultracold atomic gases.  More generally, multi-particle correlation functions have been used to experimentally characterise the fundamental properties of various systems, such as thermal Bose and Fermi gases \cite{Jeltes2007}, weakly and strongly interacting 1D Bose 
gases \cite{Kinoshita2005,Armijo2010,Fang2016}, tunnel-coupled 1D tubes \cite{Langen2015,Schweigler2016}, collision halos \cite{Perrin2007,Kheruntsyan2012,Khakimov2016}, and phenomena such as prethermalisation \cite{Gring2012} and transverse condensation \cite{RuGway2013}.  Correlations between multiple photons are also routinely used in numerous quantum optics experiments including ghost imaging \cite{Chan2009,Chen2010}, defining criteria for non-classicality \cite{Vogel2008,Ding2015}, analysing entangled states generated by parametric down conversion \cite{Bussieres2008} and characterising single photon sources \cite{URen2005}.

\begin{figure*}[tp!]
  \includegraphics[width=0.9\textwidth,]{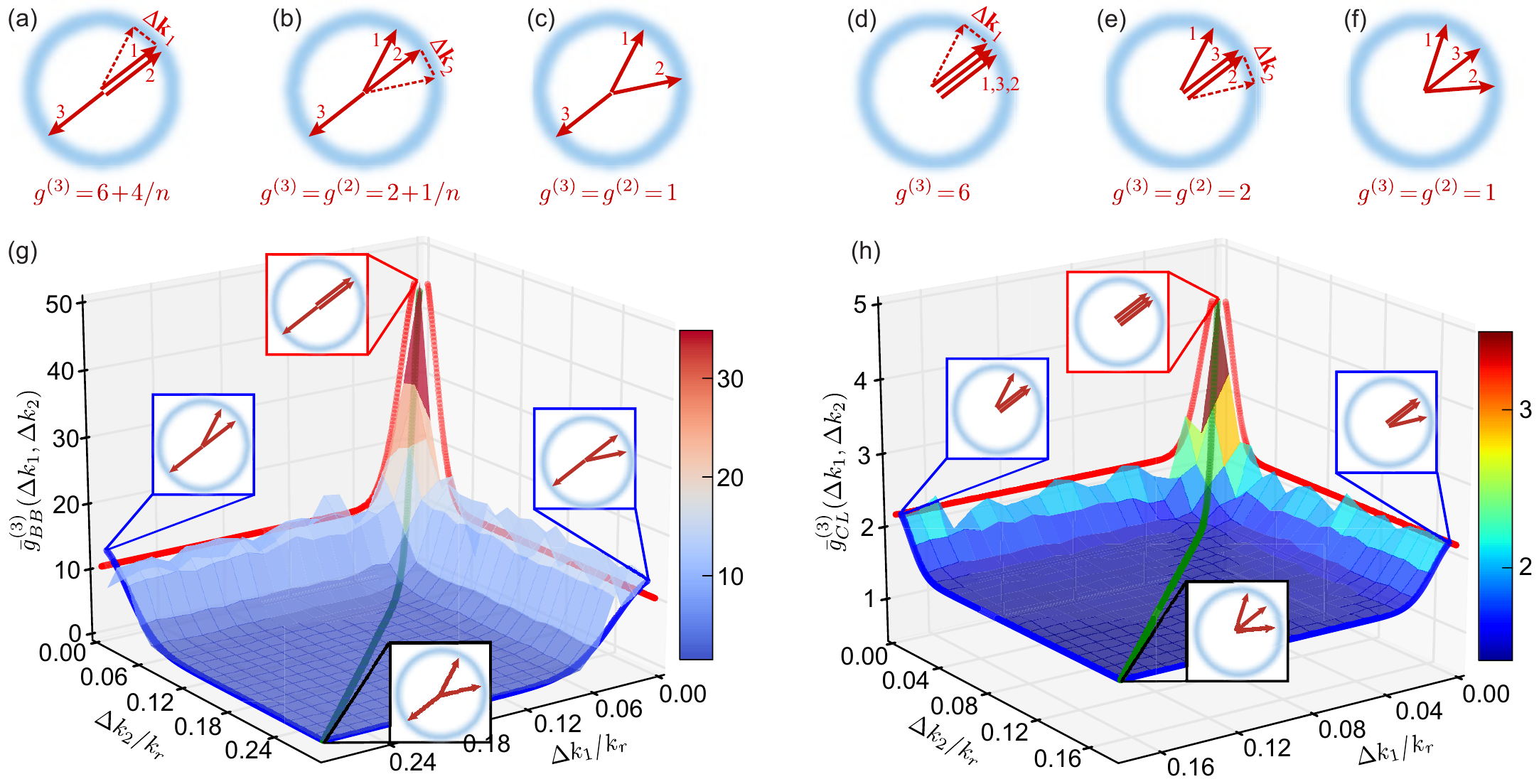}
	\caption{ \textbf{Three-body momentum correlation functions.} The various back-to-back (a-c) and collinear (d-f) correlation functions between three atoms in the scattering halo, with the maximum expected value of each correlation function indicated.  (g) Surface plot showing the correlation function between two collinear and one back-to-back atom $\bar{g}^{(3)}_{BB}(\Delta k_1,\Delta k_2)$ is shown for the $s$-wave halo that has a mean occupation of $n=0.010(5)$ atoms per mode. The red and blue solid lines show cases (a) and (b) respectively \cite{SOMs}, while the green solid line is the 1D Gaussian fit used to extract $\bar{g}^{(3)}_{BB}(0,0)$.  (h)  The correlation function between three collinear atoms $\bar{g}^{(3)}_{CL}(\Delta k_1,\Delta k_2)$ is shown for the $s$-wave halo with $n=0.44(2)$.  The red and blue solid lines show cases (d) and (e) respectively, while the green solid line is the 1D Gaussian fit used to extract $\bar{g}^{(3)}_{CL}(0,0)$ \cite{SOMs}.  Insets show visual representations of the relevant cases at four selected points on each plot.	}
	\label{fig:g3_surf}
\end{figure*}

Here, we demonstrate an experimental solution of the many-body problem as outlined above by measuring second- and third-order correlations between momentum-correlated atoms in a collisional halo between two Bose-Einstein condensates (BECs).  
The halo is generated by spontaneous $s$-wave scattering of two colliding BECs \cite{Perrin2007,Khakimov2016,SOMs}, creating a spherical  shell of pair-correlated atoms (see Fig.~\ref{fig:schematic}).  After a time-of-flight expansion, we detect the positions of individual atoms, which are mapped back to the initial momenta of the atoms directly after the collision \cite{ballistic_note}. This means that we reconstruct momentum correlation functions from the momenta of individual atoms with full 3D resolution. Thus our detector setup can be regarded as a \textit{quantum many-body momentum microscope}, complementary 
to the quantum gas \textit{in situ} microscopes created using optical lattices \cite{Bakr2009,Sherson2010,Cheneau2012,Cheuk2015,Haller2015,Omran2015} or arrays of optical tweezers \cite{Lester2015}.  
We characterise and compare all possible back-to-back and collinear atomic correlation functions for two and three atoms, showing the relationship between the different correlation functions and demonstrating a record violation of the classical Cauchy-Schwarz inequality between the peak values of correlation functions.  

The experiments start with a BEC of $\sim10^6$ ${}^4$He$^*$ atoms magnetically trapped in the $m_J=+1$ sublevel of the long lived metastable (${2}^3$S$_1$) state \cite{Vassen2012}.  The $s$-wave scattering halos of correlated pairs are produced \textit{via} a two-step process.  First, we transfer $\sim$ 95 $\%$ of the BEC atoms to the untrapped $m_J=0$ sublevel with a Raman pulse, giving the atoms a downward (i.e. along the $\mathbf{\hat{z}}$ direction) momentum of $\mathbf{K}=-\sqrt{2} k_0 \mathbf{\hat{z}}$ \cite{SOMs} in wave-number units, where $k_0 = 2\pi/\lambda$ and $\lambda=1083.2$ nm is the wavelength of a diffraction photon.  The untrapped BEC is then diffracted using a second pulse into two or more diffraction orders, using either Bragg or Kapitza-Dirac diffraction \cite{SOMs}.  Adjacent pairs of diffracted condensates then collide, producing spherical halos of spontaneously scattered atom pairs via $s$-wave collisions \cite{Perrin2007}.  Each halo has a radius in momentum space of $k_{r}\approx k_0/\sqrt{2}$ and a radial Gaussian width of $w\approx 0.03k_r$.  The average mode occupancy in each halo ranges from $n = 0.0017(17)$ to $n = 0.44(2)$.  As the scattering in our experiment is always in the spontaneous pair-production regime \cite{SOMs}, the scattering halo can be approximated by an overall quantum many-body state that is the product of independent two-mode squeezed vacuum states 
analogous to those produced by parametric down-conversion in quantum optics.   

The expanding halos then fall $\sim\!850$\!~mm (time of flight [TOF] $\sim\!416\!$~ms) onto a multi-channel plate and delay-line detector.  Due to the $19.8$~eV internal energy of the ${2}^3$S$_1$ state, the individual positions of atoms can be reconstructed in 3D, with a spatial resolution of $\sim\!~\!120$~$\mu$m in $x,y$ (momentum resolution $\sim 0.0044 k_r$), and a temporal resolution along $z$ of $\sim \!2$~ns ($\equiv 8$~nm or $3\times 10^{-7} k_r$).  As we are interested in correlations between atoms in different momentum modes, we convert position and time to momentum centered on each halo \cite{SOMs}.

\begin{figure*}[t!]
\includegraphics[width=0.93\textwidth, keepaspectratio=true]{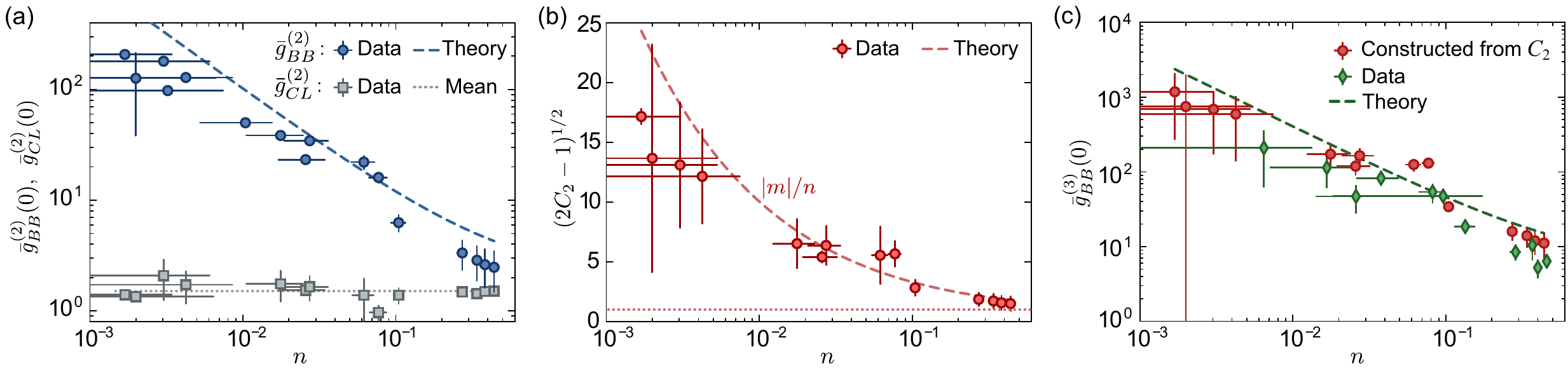}

\caption{ {\bf Peak two- and three-atom correlation amplitudes and anomalous occupancy $|m|$ vs halo mode occupancy $n$}. (a) The measured peak two-atom back-to-back and collinear correlation amplitudes $\bar{g}^{(2)}_{BB}(0)$ (blue circles) and $\bar{g}^{(2)}_{CL}(0)$ (grey squares), respectively, plotted against the average halo mode occupancy ($n$) for different halos. The dashed (blue) line shows the analytic prediction of Eq.\,(\ref{eq:g2_vs_n}).  We expect theoretically that $\bar{g}^{(2)}_{CL}(0)\!=\!2$ for all $n$, but due to the finite resolution of the detector and the bins used to calculate the correlation function, this is reduced slightly. The dotted line shows the mean value of $\bar{g}^{(2)}_{CL}(0)$. (b)~The quantity $(2C_2-1)^{1/2}\!\simeq |m|/n$ (where $C_2\!\equiv\!\bar{g}_{BB}^{(2)}(0)/\bar{g}_{CL}^{(2)}(0)$), with $|m|$ the anomalous occupancy, is plotted against $n$ along with the theoretical prediction (dashed red line) of $|m|/n\!=\!(1+1/n)^{1/2}$. The horizontal dotted line at unity is drawn for reference, showing that $|m|\!>\!n$ for all our data points. (c) $\bar{g}^{(3)}_{BB}(0,0)$ vs $n$, with green diamonds showing experimental data (extracted from fits to $\bar{g}^{(3)}_{BB}(\Delta k,\Delta k)$, as shown by the green line in Fig. \ref{fig:g3_surf}(g)) and the dashed line showing the theoretical prediction of Eq.\,(\ref{eq:g3_vs_n}). Red circles are reconstructed using experimental data for $C_2$.~  Error bars for all three plots show the combined statistical and fit uncertainties \cite{SOMs}.    }
	\label{fig:g2_vs_n}
\end{figure*}

From the reconstructed momentum for each atom, we construct various momentum correlation functions from coincidence counts between atoms within each experimental run that are averaged over all runs \cite{SOMs}.  First, we look at momentum correlations between three atoms with momenta $\mathbf{k}_3$, $\mathbf{k}_1= -\mathbf{k}_3 + \Delta \mathbf{k}_1$ and $\mathbf{k}_2= -\mathbf{k}_3 + \Delta \mathbf{k}_2$, i.e.\,the second two atoms on the opposite side of the halo to the first  [see Fig.~\ref{fig:g3_surf}(a-c) for illustration]. We define the relevant correlation function as $\bar{g}_{BB}^{(3)}(\Delta k_1,\Delta k_2)$ and refer to it as back-to-back (BB), which is averaged with respect to $\mathbf{k}_3$ over the halo and spherically integrated with respect to the directions of vectors $\Delta \mathbf{k}_1$ and $\Delta \mathbf{k}_2$.  Thus it is a function of the absolute values $\Delta k_1=|\Delta \mathbf{k}_1|$ and $\Delta k_2=|\Delta \mathbf{k}_2|$ \cite{SOMs}.

Fig.~\ref{fig:g3_surf}\,(g) shows a typical surface plot of $\bar{g}_{BB}^{(3)}(\Delta k_1,\Delta k_2)$ for the $s$-wave halo generated by Kapitza-Dirac orders $l\!=\!(-2,-3)$.  This surface plot also contains other many-body correlation functions within it, shown schematically in Figs.~\ref{fig:g3_surf}\,(a)-(c).  When $\Delta k_2\!=\!0$, we can plot $\bar{g}_{BB}^{(3)}(\Delta k_1,0)$ (red line in Fig.~\ref{fig:g3_surf}\,(g)), which will asymptotically approach $\bar{g}_{BB}^{(2)}(0)$---the two-particle correlation function with one atom on each side of the halo---for $\Delta k_1\!\gg \!\sigma_{BB}$, where $\sigma_{BB}$ is the two-particle back-to-back correlation length.  Taking $\Delta k_2\!\gg \!\sigma_{BB}$, we can also plot $\bar{g}_{BB}^{(3)}(\Delta k_1, \Delta k_2 \!\gg \!\sigma_{BB})$ (blue line), which is equivalent to $\bar{g}_{BB}^{(2)}(\Delta k_1)$ and approaches the uncorrelated case of $\bar{g}_{BB}^{(2)}(\Delta k_1\!\gg \!\sigma_{BB})=1$ for large values of $\Delta k_1$ (see \cite{SOMs} for a full discussion of the relationship between various correlation functions).  

We also measure the collinear (CL) three-atom correlation function [shown in Fig.~\ref{fig:g3_surf} (d-f)], defined analogously as $\bar{g}_{CL}^{(3)}(\Delta k_1,\Delta k_2)$, where now $\Delta k_1 \!=\!|\mathbf{k}_3-\mathbf{k}_1|$ and $\Delta k_2 \!=\!|\mathbf{k}_3-\mathbf{k}_2|$.
A surface plot of this function, measured for the Bragg halo with maximum mode occupancy, is shown in Fig.~\ref{fig:g3_surf}\,(h).  Like $\bar{g}_{BB}^{(3)}(\Delta k_1,\Delta k_2)$, this full correlation function also contains other many-body correlations: for example, $\bar{g}_{CL}^{(3)}(\Delta k_1,0)$ [red line in Fig.~\ref{fig:g3_surf}\,(h)] will asymptotically approach $\bar{g}_{CL}^{(2)}(0)$ (the two-atom collinear correlation function), while $\bar{g}_{CL}^{(3)}(\Delta k_1,\Delta k_2 \!\gg \!\sigma_{CL})$ (blue line) will yield $\bar{g}_{CL}^{(2)}(\Delta k_1)$ \cite{SOMs}.  Figs.~\ref{fig:g3_surf}\,(g) and (h) therefore show a full characterisation of the hierarchy of all three-body and two-body correlation functions present in our system.

Our collisional halo is an example of a quantum many-body system which, in the spontaneous scattering regime, satisfies Wick's factorisation scheme \cite{SOMs}.  This requires knowledge of both the normal and anomalous second-order operator moments in momentum space, $n_{\mathbf{k},\mathbf{k}+\Delta \mathbf{k}} \!=\!\langle \hat{a}^{\dagger}_{\mathbf{k}} \hat{a}_{\mathbf{k}+\Delta \mathbf{k}} \rangle$ and $m_{\mathbf{k},-\mathbf{k}+\Delta \mathbf{k}}\!=\!\langle \hat{a}_{\mathbf{k}} \hat{a}_{\mathbf{-k}+\Delta \mathbf{k}} \rangle$, with $\hat{a}^{\dagger}_{\mathbf{k}}$ and $\hat{a}_{\mathbf{k}}$ being the respective mode creation and annihilation operators, 
and the diagonal element of $n_{\mathbf{k},\mathbf{k}+\Delta \mathbf{k}}$ giving the average mode occupancy $n_{\mathbf{k}}\equiv n_{\mathbf{k},\mathbf{k}}$. Knowledge of these quantities is sufficient to reconstruct all higher-order correlation functions and thus completely solve the many-body problem for our system. 
Here, the anomalous occupancy $m_{\mathbf{k}}\equiv m_{\mathbf{k},-\mathbf{k}}$ (related to the anomalous Green's function in quantum field theory) describes the pairing field amplitude between atoms with equal but  opposite momenta, 
and is similar to the expectation value of the Cooper pair operator in the Bardeen-Cooper-Schrieffer theory of superconductivity, although in our case the pairing is between two identical bosons.

To examine these factorization properties further, we analyse the dependence of peak correlation amplitudes on the peak halo mode occupancy $n_{\mathbf{k}_0}$ and compare them with theoretical predictions.  The theory relies on the relationship between the peak anomalous occupancy $|m_{\mathbf{k}_0}|$ and $n_{\mathbf{k}_0}$: $|m_{\mathbf{k}_0}|^2=n_{\mathbf{k}_0}(n_{\mathbf{k}_0}+1)$ \cite{SOMs}.
In Fig.~\ref{fig:g2_vs_n}\,(a), we plot the measured peak back-to-back correlation amplitude between two-atoms $\bar{g}_{BB}^{(2)}(0)$, for values of average mode occupancy $n\simeq n_{\mathbf{k}_0}$ that span more than two orders of magnitude. $\bar{g}_{BB}^{(2)}(0)$ is extracted by fitting $\bar{g}_{BB}^{(2)}(\Delta k)$ with a Gaussian (for details and plots, see \cite{SOMs}). From analytic theory we expect $\bar{g}_{BB}^{(2)}(0)$ to scale with $n$ as \cite{SOMs}:
\begin{equation}\label{eq:g2_vs_n}
	\bar{g}_{BB}^{(2)}(0) = (n_{\mathbf{k}_0}^2+|m_{\mathbf{k}_0}|^2)/n_{\mathbf{k}_0}^2 \simeq 2 +1/n. 
\end{equation}
This relation is plotted as the dashed line in Fig.~\ref{fig:g2_vs_n}\,(a), which matches the data well considering that it is a no free parameters fit. For comparison we also plot the peak collinear correlation between two-atoms, $\bar{g}_{CL}^{(2)}(0)$, shown by squares in Fig.~\ref{fig:g2_vs_n}\,(a) and extracted from $\bar{g}_{CL}^{(2)}(\Delta k)$ in the same way as $\bar{g}_{BB}^{(2)}(0)$. We see values of $\bar{g}_{CL}^{(2)}(0) \simeq 1.5$, seemingly independent of the mode occupancy.  This trend is expected theoretically, although in the limit of perfect resolution we would expect $\bar{g}_{CL}^{(2)}(0)=2$ (as in the Hanbury Brown--Twiss effect \cite{Perrin2007,Perrin2008}) for all values of $n$.

From the measured $\bar{g}_{BB}^{(2)}(0)$ and $\bar{g}_{CL}^{(2)}(0)$ at each $n$ we are able to extract the key nontrivial component of all higher-order correlations in the scattering halo---the absolute value of the average anomalous occupancy $|m|\simeq |m_{\mathbf{k}_0}|$. This is found through the relation $|m_{\mathbf{k}_0}|^2/n_{\mathbf{k}_0}^2\!=\!2C_2-1$, where $C_2\!\equiv\!\bar{g}_{BB}^{(2)}(0)/\bar{g}_{CL}^{(2)}(0)$. Using the data of Fig.~\ref{fig:g2_vs_n}\,(a) to calculate $C_2$, we plot $(2C_2-1)^{1/2}\simeq |m|/n$ in Fig.~\ref{fig:g2_vs_n}\,(b).  A value of $|m|>n$ is necessary for any system to exhibit non-classical (quantum) behaviour, such as two-mode quadrature squeezing, Einstein-Podolsky-Rosen quadrature-entanglement \cite{Kheruntsyan-EPR-2005}, and Bell inequality violation \cite{Lewis-Swan2015}.  The fact that we measure values of $|m|/n\!>\!1$ for all $n$ (with $|m|/n\gg 1$ for smallest $n$) is a further demonstration of the strong quantum nature of our system.  Since all order correlation functions for this system can be expressed as a function of $n$ and $|m|$, measuring these parameters is essentially equivalent to solving the many-body problem for the collisional halo.

Following a similar analysis for the peak three-atom back-to-back correlation amplitude \cite{SOMs}, extracting $\bar{g}_{BB}^{(3)}(0,0)$ from Gaussian fits to $\bar{g}_{BB}^{(3)}(\Delta k,\Delta k)$, we plot these peak values as a function of $n$ in Fig.~\ref{fig:g2_vs_n}\,(c). Theoretically, we expect $\bar{g}_{BB}^{(3)}(0,0)$ to scale with $|m|$ and $n$ as \cite{SOMs}
\begin{equation}
	\bar{g}_{BB}^{(3)}(0,0) = (2n_{\mathbf{k}_0}^3+4n_{\mathbf{k}_0}|m_{\mathbf{k}_0}|^2)/n_{\mathbf{k}_0}^3 \simeq 6 +4/n. 
    \label{eq:g3_vs_n}
\end{equation}
This reflects the enhancement in the correlation amplitude due to both the back-to-back and collinear correlations \cite{SOMs}.  In Fig.~\ref{fig:g2_vs_n}\,(c) we plot Eq.~(\ref{eq:g3_vs_n}) as a dashed green line, which agrees quite well with the experimental data.  

Additionally, we can construct $\bar{g}_{BB}^{(3)}(0,0)$ from our measured values of $C_2$, through the relation $\bar{g}_{BB}^{(3)}(0,0)=8C_2-2$.  We plot these values in Fig. \ref{fig:g2_vs_n}\,(c), which match the theory well.  This is a direct demonstration of how lower-order correlation functions can be used to construct higher-order correlation functions, showing that measuring a finite number of correlation functions can be operationally equivalent to solving the many-body problem.

The low probability associated with four or more atom coincidence events means that we are unable to perform a full, quantitative analysis of the hierarchy of fourth- and higher-order correlation functions.  However, we are able to measure the back-to-back correlation function $\bar{g}_{BB}^{(4)}(\Delta k_1,\Delta k_2,\Delta k_3)$ for four atoms, two on each opposite side of the halo \cite{SOMs}, for $n\!=\!0.31(12)$.~This yields $\bar{g}_{BB}^{(4)}(0,0,0) \!= \!70(40)$, compared to the theoretically expected value of $\bar{g}_{BB}^{(4)}(0,0,0) \!\simeq \!24+24/n+4/n^2 \!\simeq \!143$ for this mode occupancy \cite{SOMs}.

An important feature of our BEC collision experiments compared to previous work \cite{Perrin2007,Kheruntsyan2012} is that we are able to explore a much larger parameter space, including relatively low values of $n$ and small correlation lengths \cite{SOMs}. Because of this, the values of $\bar{g}_{BB}^{(2)}(0)$ that we measure greatly exceed the maximum possible collinear correlation value of $\bar{g}_{CL}^{(2)}(0)\!=\!2$. Thus our results are the first measurements in the regime $\bar{g}_{BB}^{(2)}(0) \!\gg \!\bar{g}_{CL}^{(2)}(0)$.  This is a violation of the simplest formulation of the Cauchy-Schwarz inequality \cite{Kheruntsyan2012} for our system, which dictates that classically we would be restricted to
$\bar{g}_{BB}^{(2)}(0) \leq  \bar{g}_{CL}^{(2)}(0)$.
All previous similar measurements with ultracold atoms were limited to peak correlation amplitudes $\bar{g}_{BB}^{(2)}(0) \!\simeq \!\bar{g}_{CL}^{(2)}(0)$ \cite{Perrin2007,Kheruntsyan2012}.  This meant that they were only able to show a violation of the Cauchy-Schwarz inequality for volume-integrated atom numbers, rather than bare peak correlations \cite{SOMs}.  Therefore our measurement of $\bar{g}_{BB}^{(2)}(0) \!\gg \!\bar{g}_{CL}^{(2)}(0)$, with $C_2\!=\!\bar{g}_{BB}^{(2)}(0)/\bar{g}_{CL}^{(2)}(0)\!>\!100$, represents a more straightforward and much stronger violation of the Cauchy-Schwarz inequality (cf. the maximum value of  the corresponding correlation coefficient $C_2\!\simeq \!1.2$ measured in Ref. \cite{Kheruntsyan2012}).  In fact, to the best of our knowledge even for optical experiments the largest value measured is $C_2\!\simeq \!58$ \cite{Lee2006,SOMs}, meaning that our result of $C_2\!>\!100$ is a record for any source. 

The Cauchy-Schwarz inequality can also be formulated for higher-order correlation functions.  For three-atom correlations in our system it states 
$\bar{g}_{BB}^{(3)}(0,0) \leq  \left(\bar{g}_{CL}^{(2)}(0)\right)^{3/2}$ \cite{Ding2015}.
Again, we violate this inequality for all data in Fig.~\ref{fig:g2_vs_n}, with a maximum violation of $\simeq\! 100$.

To summarise, we have used a quantum many-body momentum microscope to analyse the spontaneous $s$-wave scattering halos of correlated atom pairs with a range of halo mode occupancies $n$ spanning over two orders of magnitude.  We measured the third-order correlation functions $\bar{g}^{(3)}_{BB}$ and $\bar{g}^{(3)}_{CL}$ and confirmed the non-trivial many-body nature of the correlations present. 
Unlike previous similar measurements, we were able to extract the absolute value of the anomalous occupancy $|m|$ as a function of $n$. $|m|$ and $n$ are all that is required for understanding and predicting all higher-order correlation functions in this system, hence solving the quantum many-body problem in this case.
We have also demonstrated a high degree of violation of the classical Cauchy-Schwarz inequality for both two and three atom correlations.  This is the first measurement for three atoms, while our two atom result beats the only previous experiment with atoms \cite{Kheruntsyan2012} by nearly two orders of magnitude. 

This demonstrated ability to measure
higher-order quantum correlations in a complex many-body system (an $s$-wave scattering halo) means that a momentum microscope will be a valuable tool for probing other many-body effects in quantum simulators that possess non-trivial correlations (although this may require additional considerations \cite{SOMs}).  Such effects
include many-body localisation and glassy dynamics \cite{He2012}, 
unconventional superconductivity \cite{Rey2009}, 
universal three-body recombination and Efimov resonances \cite{Kunitski2015}.  Other possible applications include the use of such a microscope as a direct dynamical probe of non-equilibrium many-body effects in TOF expansion. 

\vspace{5 mm}

We thank I. Bloch, C. Regal and A.-M. Rey for helpful comments and B. Henson and D. Shin for technical assistance. This work was supported through Australian Research Council (ARC) Discovery Project grants DP120101390, DP140101763 and DP160102337. S.S.H. is supported by ARC Discovery Early Career Research Award DE150100315.  A.G.T. is supported by ARC Future Fellowship grant FT100100468.

\bibliography{Refs}

\begin{thebibliography}{62}%
\makeatletter
\providecommand \@ifxundefined [1]{%
 \@ifx{#1\undefined}
}%
\providecommand \@ifnum [1]{%
 \ifnum #1\expandafter \@firstoftwo
 \else \expandafter \@secondoftwo
 \fi
}%
\providecommand \@ifx [1]{%
 \ifx #1\expandafter \@firstoftwo
 \else \expandafter \@secondoftwo
 \fi
}%
\providecommand \natexlab [1]{#1}%
\providecommand \enquote  [1]{``#1''}%
\providecommand \bibnamefont  [1]{#1}%
\providecommand \bibfnamefont [1]{#1}%
\providecommand \citenamefont [1]{#1}%
\providecommand \href@noop [0]{\@secondoftwo}%
\providecommand \href [0]{\begingroup \@sanitize@url \@href}%
\providecommand \@href[1]{\@@startlink{#1}\@@href}%
\providecommand \@@href[1]{\endgroup#1\@@endlink}%
\providecommand \@sanitize@url [0]{\catcode `\\12\catcode `\$12\catcode
  `\&12\catcode `\#12\catcode `\^12\catcode `\_12\catcode `\%12\relax}%
\providecommand \@@startlink[1]{}%
\providecommand \@@endlink[0]{}%
\providecommand \url  [0]{\begingroup\@sanitize@url \@url }%
\providecommand \@url [1]{\endgroup\@href {#1}{\urlprefix }}%
\providecommand \urlprefix  [0]{URL }%
\providecommand \Eprint [0]{\href }%
\providecommand \doibase [0]{http://dx.doi.org/}%
\providecommand \selectlanguage [0]{\@gobble}%
\providecommand \bibinfo  [0]{\@secondoftwo}%
\providecommand \bibfield  [0]{\@secondoftwo}%
\providecommand \translation [1]{[#1]}%
\providecommand \BibitemOpen [0]{}%
\providecommand \bibitemStop [0]{}%
\providecommand \bibitemNoStop [0]{.\EOS\space}%
\providecommand \EOS [0]{\spacefactor3000\relax}%
\providecommand \BibitemShut  [1]{\csname bibitem#1\endcsname}%
\let\auto@bib@innerbib\@empty
\bibitem [{\citenamefont {Schweigler}\ \emph {et~al.}()\citenamefont
  {Schweigler}, \citenamefont {Kasper}, \citenamefont {Erne}, \citenamefont
  {Rauer}, \citenamefont {Langen}, \citenamefont {Gasenzer}, \citenamefont
  {Berges},\ and\ \citenamefont {Schmiedmayer}}]{Schweigler2016}%
  \BibitemOpen
  \bibfield  {author} {\bibinfo {author} {\bibfnamefont {T.}~\bibnamefont
  {Schweigler}}, \bibinfo {author} {\bibfnamefont {V.}~\bibnamefont {Kasper}},
  \bibinfo {author} {\bibfnamefont {S.}~\bibnamefont {Erne}}, \bibinfo {author}
  {\bibfnamefont {B.}~\bibnamefont {Rauer}}, \bibinfo {author} {\bibfnamefont
  {T.}~\bibnamefont {Langen}}, \bibinfo {author} {\bibfnamefont
  {T.}~\bibnamefont {Gasenzer}}, \bibinfo {author} {\bibfnamefont
  {J.}~\bibnamefont {Berges}}, \ and\ \bibinfo {author} {\bibfnamefont
  {J.}~\bibnamefont {Schmiedmayer}},\ }\href {http://arxiv.org/abs/1505.03126}
  {\bibinfo  {journal} {arXiv:1505.03126}\ }\BibitemShut {NoStop}%
\bibitem [{\citenamefont {Shavitt}\ and\ \citenamefont
  {Bartlett}(2009)}]{shavitt2009many}%
  \BibitemOpen
\bibfield  {journal} {  }\bibfield  {author} {\bibinfo {author} {\bibfnamefont
  {I.}~\bibnamefont {Shavitt}}\ and\ \bibinfo {author} {\bibfnamefont {R.~J.}\
  \bibnamefont {Bartlett}},\ }\href@noop {} {\emph {\bibinfo {title} {Many-body
  methods in chemistry and physics: MBPT and coupled-cluster theory}}}\
  (\bibinfo  {publisher} {Cambridge university press},\ \bibinfo {year}
  {2009})\BibitemShut {NoStop}%
\bibitem [{\citenamefont {Kira}\ and\ \citenamefont
  {Koch}(2011)}]{kira2011semiconductor}%
  \BibitemOpen
  \bibfield  {author} {\bibinfo {author} {\bibfnamefont {M.}~\bibnamefont
  {Kira}}\ and\ \bibinfo {author} {\bibfnamefont {S.~W.}\ \bibnamefont
  {Koch}},\ }\href@noop {} {\emph {\bibinfo {title} {Semiconductor quantum
  optics}}}\ (\bibinfo  {publisher} {Cambridge University Press},\ \bibinfo
  {year} {2011})\BibitemShut {NoStop}%
\bibitem [{\citenamefont {{R. G. Dall}}\ \emph {et~al.}(2013)\citenamefont {{R.
  G. Dall}}, \citenamefont {{A. G. Manning}}, \citenamefont {{S. S. Hodgman}},
  \citenamefont {{Wu RuGway}}, \citenamefont {{K. V. Kheruntsyan}},\ and\
  \citenamefont {{A. G. Truscott}}}]{Dall2013_idealnbody}%
  \BibitemOpen
  \bibfield  {author} {\bibinfo {author} {\bibnamefont {{R. G. Dall}}},
  \bibinfo {author} {\bibnamefont {{A. G. Manning}}}, \bibinfo {author}
  {\bibnamefont {{S. S. Hodgman}}}, \bibinfo {author} {\bibnamefont {{Wu
  RuGway}}}, \bibinfo {author} {\bibnamefont {{K. V. Kheruntsyan}}}, \ and\
  \bibinfo {author} {\bibnamefont {{A. G. Truscott}}},\ }\href {\doibase
  http://dx.doi.org/10.1038/nphys2632} {\bibfield  {journal} {\bibinfo
  {journal} {Nat. Phys.}\ }\textbf {\bibinfo {volume} {9}},\ \bibinfo {pages}
  {341} (\bibinfo {year} {2013})}\BibitemShut {NoStop}%
\bibitem [{\citenamefont {Langen}\ \emph {et~al.}(2015)\citenamefont {Langen},
  \citenamefont {Erne}, \citenamefont {Geiger}, \citenamefont {Rauer},
  \citenamefont {Schweigler}, \citenamefont {Kuhnert}, \citenamefont
  {Rohringer}, \citenamefont {Mazets}, \citenamefont {Gasenzer},\ and\
  \citenamefont {Schmiedmayer}}]{Langen2015}%
  \BibitemOpen
  \bibfield  {author} {\bibinfo {author} {\bibfnamefont {T.}~\bibnamefont
  {Langen}}, \bibinfo {author} {\bibfnamefont {S.}~\bibnamefont {Erne}},
  \bibinfo {author} {\bibfnamefont {R.}~\bibnamefont {Geiger}}, \bibinfo
  {author} {\bibfnamefont {B.}~\bibnamefont {Rauer}}, \bibinfo {author}
  {\bibfnamefont {T.}~\bibnamefont {Schweigler}}, \bibinfo {author}
  {\bibfnamefont {M.}~\bibnamefont {Kuhnert}}, \bibinfo {author} {\bibfnamefont
  {W.}~\bibnamefont {Rohringer}}, \bibinfo {author} {\bibfnamefont {I.~E.}\
  \bibnamefont {Mazets}}, \bibinfo {author} {\bibfnamefont {T.}~\bibnamefont
  {Gasenzer}}, \ and\ \bibinfo {author} {\bibfnamefont {J.}~\bibnamefont
  {Schmiedmayer}},\ }\href {\doibase 10.1126/science.1257026} {\bibfield
  {journal} {\bibinfo  {journal} {Science}\ }\textbf {\bibinfo {volume}
  {348}},\ \bibinfo {pages} {207} (\bibinfo {year} {2015})}\BibitemShut
  {NoStop}%
\bibitem [{\citenamefont {Jeltes}\ \emph {et~al.}(2007)\citenamefont {Jeltes},
  \citenamefont {McNamara}, \citenamefont {Hogervorst}, \citenamefont {Vassen},
  \citenamefont {Krachmalnicoff}, \citenamefont {Schellekens}, \citenamefont
  {Perrin}, \citenamefont {Chang}, \citenamefont {Boiron}, \citenamefont
  {Aspect},\ and\ \citenamefont {Westbrook}}]{Jeltes2007}%
  \BibitemOpen
  \bibfield  {author} {\bibinfo {author} {\bibfnamefont {T.}~\bibnamefont
  {Jeltes}}, \bibinfo {author} {\bibfnamefont {J.~M.}\ \bibnamefont
  {McNamara}}, \bibinfo {author} {\bibfnamefont {W.}~\bibnamefont
  {Hogervorst}}, \bibinfo {author} {\bibfnamefont {W.}~\bibnamefont {Vassen}},
  \bibinfo {author} {\bibfnamefont {V.}~\bibnamefont {Krachmalnicoff}},
  \bibinfo {author} {\bibfnamefont {M.}~\bibnamefont {Schellekens}}, \bibinfo
  {author} {\bibfnamefont {A.}~\bibnamefont {Perrin}}, \bibinfo {author}
  {\bibfnamefont {H.}~\bibnamefont {Chang}}, \bibinfo {author} {\bibfnamefont
  {D.}~\bibnamefont {Boiron}}, \bibinfo {author} {\bibfnamefont
  {A.}~\bibnamefont {Aspect}}, \ and\ \bibinfo {author} {\bibfnamefont {C.~I.}\
  \bibnamefont {Westbrook}},\ }\href {\doibase 10.1038/nature05513} {\bibfield
  {journal} {\bibinfo  {journal} {Nature}\ }\textbf {\bibinfo {volume} {445}},\
  \bibinfo {pages} {402} (\bibinfo {year} {2007})}\BibitemShut {NoStop}%
\bibitem [{\citenamefont {Kinoshita}\ \emph {et~al.}(2005)\citenamefont
  {Kinoshita}, \citenamefont {Wenger},\ and\ \citenamefont
  {Weiss}}]{Kinoshita2005}%
  \BibitemOpen
  \bibfield  {author} {\bibinfo {author} {\bibfnamefont {T.}~\bibnamefont
  {Kinoshita}}, \bibinfo {author} {\bibfnamefont {T.}~\bibnamefont {Wenger}}, \
  and\ \bibinfo {author} {\bibfnamefont {D.~S.}\ \bibnamefont {Weiss}},\ }\href
  {\doibase 10.1103/PhysRevLett.95.190406} {\bibfield  {journal} {\bibinfo
  {journal} {Phys. Rev. Lett.}\ }\textbf {\bibinfo {volume} {95}},\ \bibinfo
  {pages} {190406} (\bibinfo {year} {2005})}\BibitemShut {NoStop}%
\bibitem [{\citenamefont {Armijo}\ \emph {et~al.}(2010)\citenamefont {Armijo},
  \citenamefont {Jacqmin}, \citenamefont {Kheruntsyan},\ and\ \citenamefont
  {Bouchoule}}]{Armijo2010}%
  \BibitemOpen
  \bibfield  {author} {\bibinfo {author} {\bibfnamefont {J.}~\bibnamefont
  {Armijo}}, \bibinfo {author} {\bibfnamefont {T.}~\bibnamefont {Jacqmin}},
  \bibinfo {author} {\bibfnamefont {K.~V.}\ \bibnamefont {Kheruntsyan}}, \ and\
  \bibinfo {author} {\bibfnamefont {I.}~\bibnamefont {Bouchoule}},\ }\href
  {\doibase 10.1103/PhysRevLett.105.230402} {\bibfield  {journal} {\bibinfo
  {journal} {Phys. Rev. Lett.}\ }\textbf {\bibinfo {volume} {105}},\ \bibinfo
  {pages} {230402} (\bibinfo {year} {2010})}\BibitemShut {NoStop}%
\bibitem [{\citenamefont {Fang}\ \emph {et~al.}(2016)\citenamefont {Fang},
  \citenamefont {Johnson}, \citenamefont {Roscilde},\ and\ \citenamefont
  {Bouchoule}}]{Fang2016}%
  \BibitemOpen
  \bibfield  {author} {\bibinfo {author} {\bibfnamefont {B.}~\bibnamefont
  {Fang}}, \bibinfo {author} {\bibfnamefont {A.}~\bibnamefont {Johnson}},
  \bibinfo {author} {\bibfnamefont {T.}~\bibnamefont {Roscilde}}, \ and\
  \bibinfo {author} {\bibfnamefont {I.}~\bibnamefont {Bouchoule}},\ }\href
  {\doibase 10.1103/PhysRevLett.116.050402} {\bibfield  {journal} {\bibinfo
  {journal} {Phys. Rev. Lett.}\ }\textbf {\bibinfo {volume} {116}},\ \bibinfo
  {pages} {050402} (\bibinfo {year} {2016})}\BibitemShut {NoStop}%
\bibitem [{\citenamefont {Perrin}\ \emph {et~al.}(2007)\citenamefont {Perrin},
  \citenamefont {Chang}, \citenamefont {Krachmalnicoff}, \citenamefont
  {Schellekens}, \citenamefont {Boiron}, \citenamefont {Aspect},\ and\
  \citenamefont {Westbrook}}]{Perrin2007}%
  \BibitemOpen
  \bibfield  {author} {\bibinfo {author} {\bibfnamefont {A.}~\bibnamefont
  {Perrin}}, \bibinfo {author} {\bibfnamefont {H.}~\bibnamefont {Chang}},
  \bibinfo {author} {\bibfnamefont {V.}~\bibnamefont {Krachmalnicoff}},
  \bibinfo {author} {\bibfnamefont {M.}~\bibnamefont {Schellekens}}, \bibinfo
  {author} {\bibfnamefont {D.}~\bibnamefont {Boiron}}, \bibinfo {author}
  {\bibfnamefont {A.}~\bibnamefont {Aspect}}, \ and\ \bibinfo {author}
  {\bibfnamefont {C.~I.}\ \bibnamefont {Westbrook}},\ }\href {\doibase
  10.1103/PhysRevLett.99.150405} {\bibfield  {journal} {\bibinfo  {journal}
  {Phys. Rev. Lett.}\ }\textbf {\bibinfo {volume} {99}},\ \bibinfo {pages}
  {150405} (\bibinfo {year} {2007})}\BibitemShut {NoStop}%
\bibitem [{\citenamefont {Kheruntsyan}\ \emph {et~al.}(2012)\citenamefont
  {Kheruntsyan}, \citenamefont {Jaskula}, \citenamefont {Deuar}, \citenamefont
  {Bonneau}, \citenamefont {Partridge}, \citenamefont {Ruaudel}, \citenamefont
  {Lopes}, \citenamefont {Boiron},\ and\ \citenamefont
  {Westbrook}}]{Kheruntsyan2012}%
  \BibitemOpen
  \bibfield  {author} {\bibinfo {author} {\bibfnamefont {K.~V.}\ \bibnamefont
  {Kheruntsyan}}, \bibinfo {author} {\bibfnamefont {J.-C.}\ \bibnamefont
  {Jaskula}}, \bibinfo {author} {\bibfnamefont {P.}~\bibnamefont {Deuar}},
  \bibinfo {author} {\bibfnamefont {M.}~\bibnamefont {Bonneau}}, \bibinfo
  {author} {\bibfnamefont {G.~B.}\ \bibnamefont {Partridge}}, \bibinfo {author}
  {\bibfnamefont {J.}~\bibnamefont {Ruaudel}}, \bibinfo {author} {\bibfnamefont
  {R.}~\bibnamefont {Lopes}}, \bibinfo {author} {\bibfnamefont
  {D.}~\bibnamefont {Boiron}}, \ and\ \bibinfo {author} {\bibfnamefont {C.~I.}\
  \bibnamefont {Westbrook}},\ }\href {\doibase 10.1103/PhysRevLett.108.260401}
  {\bibfield  {journal} {\bibinfo  {journal} {Phys. Rev. Lett.}\ }\textbf
  {\bibinfo {volume} {108}},\ \bibinfo {pages} {260401} (\bibinfo {year}
  {2012})}\BibitemShut {NoStop}%
\bibitem [{\citenamefont {{R. I. Khakimov}}\ \emph {et~al.}(2016)\citenamefont
  {{R. I. Khakimov}}, \citenamefont {{B. M. Henson}}, \citenamefont {{D. K.
  Shin}}, \citenamefont {{S. S. Hodgman}}, \citenamefont {{R. G. Dall}},
  \citenamefont {{K. G. H. Baldwin}},\ and\ \citenamefont {{A. G.
  Truscott}}}]{Khakimov2016}%
  \BibitemOpen
  \bibfield  {author} {\bibinfo {author} {\bibnamefont {{R. I. Khakimov}}},
  \bibinfo {author} {\bibnamefont {{B. M. Henson}}}, \bibinfo {author}
  {\bibnamefont {{D. K. Shin}}}, \bibinfo {author} {\bibnamefont {{S. S.
  Hodgman}}}, \bibinfo {author} {\bibnamefont {{R. G. Dall}}}, \bibinfo
  {author} {\bibnamefont {{K. G. H. Baldwin}}}, \ and\ \bibinfo {author}
  {\bibnamefont {{A. G. Truscott}}},\ }\href {\doibase 10.1038/nature20154}
  {\bibfield  {journal} {\bibinfo  {journal} {Nature}\ }\textbf {\bibinfo
  {volume} {540}},\ \bibinfo {pages} {100} (\bibinfo {year}
  {2016})}\BibitemShut {NoStop}%
\bibitem [{\citenamefont {Gring}\ \emph {et~al.}(2012)\citenamefont {Gring},
  \citenamefont {Kuhnert}, \citenamefont {Langen}, \citenamefont {Kitagawa},
  \citenamefont {Rauer}, \citenamefont {Schreitl}, \citenamefont {Mazets},
  \citenamefont {Smith}, \citenamefont {Demler},\ and\ \citenamefont
  {Schmiedmayer}}]{Gring2012}%
  \BibitemOpen
  \bibfield  {author} {\bibinfo {author} {\bibfnamefont {M.}~\bibnamefont
  {Gring}}, \bibinfo {author} {\bibfnamefont {M.}~\bibnamefont {Kuhnert}},
  \bibinfo {author} {\bibfnamefont {T.}~\bibnamefont {Langen}}, \bibinfo
  {author} {\bibfnamefont {T.}~\bibnamefont {Kitagawa}}, \bibinfo {author}
  {\bibfnamefont {B.}~\bibnamefont {Rauer}}, \bibinfo {author} {\bibfnamefont
  {M.}~\bibnamefont {Schreitl}}, \bibinfo {author} {\bibfnamefont
  {I.}~\bibnamefont {Mazets}}, \bibinfo {author} {\bibfnamefont {D.~A.}\
  \bibnamefont {Smith}}, \bibinfo {author} {\bibfnamefont {E.}~\bibnamefont
  {Demler}}, \ and\ \bibinfo {author} {\bibfnamefont {J.}~\bibnamefont
  {Schmiedmayer}},\ }\href {\doibase 10.1126/science.1224953} {\bibfield
  {journal} {\bibinfo  {journal} {Science}\ }\textbf {\bibinfo {volume}
  {337}},\ \bibinfo {pages} {1318} (\bibinfo {year} {2012})}\BibitemShut
  {NoStop}%
\bibitem [{\citenamefont {RuGway}\ \emph {et~al.}(2013)\citenamefont {RuGway},
  \citenamefont {Manning}, \citenamefont {Hodgman}, \citenamefont {Dall},
  \citenamefont {Truscott}, \citenamefont {Lamberton},\ and\ \citenamefont
  {Kheruntsyan}}]{RuGway2013}%
  \BibitemOpen
  \bibfield  {author} {\bibinfo {author} {\bibfnamefont {W.}~\bibnamefont
  {RuGway}}, \bibinfo {author} {\bibfnamefont {A.~G.}\ \bibnamefont {Manning}},
  \bibinfo {author} {\bibfnamefont {S.~S.}\ \bibnamefont {Hodgman}}, \bibinfo
  {author} {\bibfnamefont {R.~G.}\ \bibnamefont {Dall}}, \bibinfo {author}
  {\bibfnamefont {A.~G.}\ \bibnamefont {Truscott}}, \bibinfo {author}
  {\bibfnamefont {T.}~\bibnamefont {Lamberton}}, \ and\ \bibinfo {author}
  {\bibfnamefont {K.~V.}\ \bibnamefont {Kheruntsyan}},\ }\href {\doibase
  10.1103/PhysRevLett.111.093601} {\bibfield  {journal} {\bibinfo  {journal}
  {Phys. Rev. Lett.}\ }\textbf {\bibinfo {volume} {111}},\ \bibinfo {pages}
  {093601} (\bibinfo {year} {2013})}\BibitemShut {NoStop}%
\bibitem [{\citenamefont {Chan}\ \emph {et~al.}(2009)\citenamefont {Chan},
  \citenamefont {O'Sullivan},\ and\ \citenamefont {Boyd}}]{Chan2009}%
  \BibitemOpen
  \bibfield  {author} {\bibinfo {author} {\bibfnamefont {K.~W.~C.}\
  \bibnamefont {Chan}}, \bibinfo {author} {\bibfnamefont {M.~N.}\ \bibnamefont
  {O'Sullivan}}, \ and\ \bibinfo {author} {\bibfnamefont {R.~W.}\ \bibnamefont
  {Boyd}},\ }\href {\doibase 10.1364/OL.34.003343} {\bibfield  {journal}
  {\bibinfo  {journal} {Opt. Lett.}\ }\textbf {\bibinfo {volume} {34}},\
  \bibinfo {pages} {3343} (\bibinfo {year} {2009})}\BibitemShut {NoStop}%
\bibitem [{\citenamefont {Chen}\ \emph {et~al.}(2010)\citenamefont {Chen},
  \citenamefont {Agafonov}, \citenamefont {Luo}, \citenamefont {Liu},
  \citenamefont {Xian}, \citenamefont {Chekhova},\ and\ \citenamefont
  {Wu}}]{Chen2010}%
  \BibitemOpen
  \bibfield  {author} {\bibinfo {author} {\bibfnamefont {X.-H.}\ \bibnamefont
  {Chen}}, \bibinfo {author} {\bibfnamefont {I.~N.}\ \bibnamefont {Agafonov}},
  \bibinfo {author} {\bibfnamefont {K.-H.}\ \bibnamefont {Luo}}, \bibinfo
  {author} {\bibfnamefont {Q.}~\bibnamefont {Liu}}, \bibinfo {author}
  {\bibfnamefont {R.}~\bibnamefont {Xian}}, \bibinfo {author} {\bibfnamefont
  {M.~V.}\ \bibnamefont {Chekhova}}, \ and\ \bibinfo {author} {\bibfnamefont
  {L.-A.}\ \bibnamefont {Wu}},\ }\href {\doibase 10.1364/OL.35.001166}
  {\bibfield  {journal} {\bibinfo  {journal} {Opt. Lett.}\ }\textbf {\bibinfo
  {volume} {35}},\ \bibinfo {pages} {1166} (\bibinfo {year}
  {2010})}\BibitemShut {NoStop}%
\bibitem [{\citenamefont {Vogel}(2008)}]{Vogel2008}%
  \BibitemOpen
  \bibfield  {author} {\bibinfo {author} {\bibfnamefont {W.}~\bibnamefont
  {Vogel}},\ }\href {\doibase 10.1103/PhysRevLett.100.013605} {\bibfield
  {journal} {\bibinfo  {journal} {Phys. Rev. Lett.}\ }\textbf {\bibinfo
  {volume} {100}},\ \bibinfo {pages} {013605} (\bibinfo {year}
  {2008})}\BibitemShut {NoStop}%
\bibitem [{\citenamefont {Ding}\ \emph {et~al.}(2015)\citenamefont {Ding},
  \citenamefont {Zhang}, \citenamefont {Shi}, \citenamefont {Zhou},
  \citenamefont {Li}, \citenamefont {Shi},\ and\ \citenamefont
  {Guo}}]{Ding2015}%
  \BibitemOpen
  \bibfield  {author} {\bibinfo {author} {\bibfnamefont {D.-S.}\ \bibnamefont
  {Ding}}, \bibinfo {author} {\bibfnamefont {W.}~\bibnamefont {Zhang}},
  \bibinfo {author} {\bibfnamefont {S.}~\bibnamefont {Shi}}, \bibinfo {author}
  {\bibfnamefont {Z.-Y.}\ \bibnamefont {Zhou}}, \bibinfo {author}
  {\bibfnamefont {Y.}~\bibnamefont {Li}}, \bibinfo {author} {\bibfnamefont
  {B.-S.}\ \bibnamefont {Shi}}, \ and\ \bibinfo {author} {\bibfnamefont
  {G.-C.}\ \bibnamefont {Guo}},\ }\href {\doibase 10.1364/OPTICA.2.000642}
  {\bibfield  {journal} {\bibinfo  {journal} {Optica}\ }\textbf {\bibinfo
  {volume} {2}},\ \bibinfo {pages} {642} (\bibinfo {year} {2015})}\BibitemShut
  {NoStop}%
\bibitem [{\citenamefont {Bussi\`{e}res}\ \emph {et~al.}(2008)\citenamefont
  {Bussi\`{e}res}, \citenamefont {Slater}, \citenamefont {Godbout},\ and\
  \citenamefont {Tittel}}]{Bussieres2008}%
  \BibitemOpen
  \bibfield  {author} {\bibinfo {author} {\bibfnamefont {F.}~\bibnamefont
  {Bussi\`{e}res}}, \bibinfo {author} {\bibfnamefont {J.~A.}\ \bibnamefont
  {Slater}}, \bibinfo {author} {\bibfnamefont {N.}~\bibnamefont {Godbout}}, \
  and\ \bibinfo {author} {\bibfnamefont {W.}~\bibnamefont {Tittel}},\ }\href
  {\doibase 10.1364/OE.16.017060} {\bibfield  {journal} {\bibinfo  {journal}
  {Opt. Express}\ }\textbf {\bibinfo {volume} {16}},\ \bibinfo {pages} {17060}
  (\bibinfo {year} {2008})}\BibitemShut {NoStop}%
\bibitem [{\citenamefont {U'Ren}\ \emph {et~al.}(2005)\citenamefont {U'Ren},
  \citenamefont {Silberhorn}, \citenamefont {Ball}, \citenamefont {Banaszek},\
  and\ \citenamefont {Walmsley}}]{URen2005}%
  \BibitemOpen
  \bibfield  {author} {\bibinfo {author} {\bibfnamefont {A.~B.}\ \bibnamefont
  {U'Ren}}, \bibinfo {author} {\bibfnamefont {C.}~\bibnamefont {Silberhorn}},
  \bibinfo {author} {\bibfnamefont {J.~L.}\ \bibnamefont {Ball}}, \bibinfo
  {author} {\bibfnamefont {K.}~\bibnamefont {Banaszek}}, \ and\ \bibinfo
  {author} {\bibfnamefont {I.~A.}\ \bibnamefont {Walmsley}},\ }\href {\doibase
  10.1103/PhysRevA.72.021802} {\bibfield  {journal} {\bibinfo  {journal} {Phys.
  Rev. A}\ }\textbf {\bibinfo {volume} {72}},\ \bibinfo {pages} {021802}
  (\bibinfo {year} {2005})}\BibitemShut {NoStop}%
\bibitem [{SOM()}]{SOMs}%
  \BibitemOpen
  \href@noop {} {}\bibinfo {note} {See Supplemental Material for
  details}\BibitemShut {NoStop}%
\bibitem [{bal()}]{ballistic_note}%
  \BibitemOpen
  \href@noop {} {}\bibinfo {note} {Since the halo density is low enough, the
  $s$-wave interactions among the halo atoms are minimal and therefore the
  expansion can be assumed to be ballistic.}\BibitemShut {Stop}%
\bibitem [{\citenamefont {Bakr}\ \emph {et~al.}(2009)\citenamefont {Bakr},
  \citenamefont {Gillen}, \citenamefont {Peng}, \citenamefont {F\"{o}lling},\
  and\ \citenamefont {Greiner}}]{Bakr2009}%
  \BibitemOpen
  \bibfield  {author} {\bibinfo {author} {\bibfnamefont {W.~S.}\ \bibnamefont
  {Bakr}}, \bibinfo {author} {\bibfnamefont {J.~I.}\ \bibnamefont {Gillen}},
  \bibinfo {author} {\bibfnamefont {A.}~\bibnamefont {Peng}}, \bibinfo {author}
  {\bibfnamefont {S.}~\bibnamefont {F\"{o}lling}}, \ and\ \bibinfo {author}
  {\bibfnamefont {M.}~\bibnamefont {Greiner}},\ }\href
  {http://www.nature.com/nature/journal/v462/n7269/full/nature08482.html}
  {\bibfield  {journal} {\bibinfo  {journal} {Nature}\ }\textbf {\bibinfo
  {volume} {462}},\ \bibinfo {pages} {74} (\bibinfo {year} {2009})}\BibitemShut
  {NoStop}%
\bibitem [{\citenamefont {Sherson}\ \emph {et~al.}(2010)\citenamefont
  {Sherson}, \citenamefont {Weitenberg}, \citenamefont {Endres}, \citenamefont
  {Cheneau}, \citenamefont {Bloch},\ and\ \citenamefont {Kuhr}}]{Sherson2010}%
  \BibitemOpen
  \bibfield  {author} {\bibinfo {author} {\bibfnamefont {J.~F.}\ \bibnamefont
  {Sherson}}, \bibinfo {author} {\bibfnamefont {C.}~\bibnamefont {Weitenberg}},
  \bibinfo {author} {\bibfnamefont {M.}~\bibnamefont {Endres}}, \bibinfo
  {author} {\bibfnamefont {M.}~\bibnamefont {Cheneau}}, \bibinfo {author}
  {\bibfnamefont {I.}~\bibnamefont {Bloch}}, \ and\ \bibinfo {author}
  {\bibfnamefont {S.}~\bibnamefont {Kuhr}},\ }\href
  {http://www.nature.com/nature/journal/v467/n7311/full/nature09378.html}
  {\bibfield  {journal} {\bibinfo  {journal} {Nature}\ }\textbf {\bibinfo
  {volume} {467}},\ \bibinfo {pages} {68} (\bibinfo {year} {2010})}\BibitemShut
  {NoStop}%
\bibitem [{\citenamefont {Cheneau}\ \emph {et~al.}(2012)\citenamefont
  {Cheneau}, \citenamefont {Barmettler}, \citenamefont {Poletti}, \citenamefont
  {Endres}, \citenamefont {Schau{\ss}}, \citenamefont {Fukuhara}, \citenamefont
  {Gross}, \citenamefont {Bloch}, \citenamefont {Kollath},\ and\ \citenamefont
  {Kuhr}}]{Cheneau2012}%
  \BibitemOpen
  \bibfield  {author} {\bibinfo {author} {\bibfnamefont {M.}~\bibnamefont
  {Cheneau}}, \bibinfo {author} {\bibfnamefont {P.}~\bibnamefont {Barmettler}},
  \bibinfo {author} {\bibfnamefont {D.}~\bibnamefont {Poletti}}, \bibinfo
  {author} {\bibfnamefont {M.}~\bibnamefont {Endres}}, \bibinfo {author}
  {\bibfnamefont {P.}~\bibnamefont {Schau{\ss}}}, \bibinfo {author}
  {\bibfnamefont {T.}~\bibnamefont {Fukuhara}}, \bibinfo {author}
  {\bibfnamefont {C.}~\bibnamefont {Gross}}, \bibinfo {author} {\bibfnamefont
  {I.}~\bibnamefont {Bloch}}, \bibinfo {author} {\bibfnamefont
  {C.}~\bibnamefont {Kollath}}, \ and\ \bibinfo {author} {\bibfnamefont
  {S.}~\bibnamefont {Kuhr}},\ }\href {\doibase 10.1038/nature10748} {\bibfield
  {journal} {\bibinfo  {journal} {Nature}\ }\textbf {\bibinfo {volume} {481}},\
  \bibinfo {pages} {484–487} (\bibinfo {year} {2012})}\BibitemShut {NoStop}%
\bibitem [{\citenamefont {Cheuk}\ \emph {et~al.}(2015)\citenamefont {Cheuk},
  \citenamefont {Nichols}, \citenamefont {Okan}, \citenamefont {Gersdorf},
  \citenamefont {Ramasesh}, \citenamefont {Bakr}, \citenamefont {Lompe},\ and\
  \citenamefont {Zwierlein}}]{Cheuk2015}%
  \BibitemOpen
  \bibfield  {author} {\bibinfo {author} {\bibfnamefont {L.~W.}\ \bibnamefont
  {Cheuk}}, \bibinfo {author} {\bibfnamefont {M.~A.}\ \bibnamefont {Nichols}},
  \bibinfo {author} {\bibfnamefont {M.}~\bibnamefont {Okan}}, \bibinfo {author}
  {\bibfnamefont {T.}~\bibnamefont {Gersdorf}}, \bibinfo {author}
  {\bibfnamefont {V.~V.}\ \bibnamefont {Ramasesh}}, \bibinfo {author}
  {\bibfnamefont {W.~S.}\ \bibnamefont {Bakr}}, \bibinfo {author}
  {\bibfnamefont {T.}~\bibnamefont {Lompe}}, \ and\ \bibinfo {author}
  {\bibfnamefont {M.~W.}\ \bibnamefont {Zwierlein}},\ }\href {\doibase
  10.1103/PhysRevLett.114.193001} {\bibfield  {journal} {\bibinfo  {journal}
  {Phys. Rev. Lett.}\ }\textbf {\bibinfo {volume} {114}},\ \bibinfo {pages}
  {193001} (\bibinfo {year} {2015})}\BibitemShut {NoStop}%
\bibitem [{\citenamefont {Haller}\ \emph {et~al.}(2015)\citenamefont {Haller},
  \citenamefont {Hudson}, \citenamefont {Kelly}, \citenamefont {Cotta},
  \citenamefont {Peaudecerf}, \citenamefont {Bruce},\ and\ \citenamefont
  {Kuhr}}]{Haller2015}%
  \BibitemOpen
  \bibfield  {author} {\bibinfo {author} {\bibfnamefont {E.}~\bibnamefont
  {Haller}}, \bibinfo {author} {\bibfnamefont {J.}~\bibnamefont {Hudson}},
  \bibinfo {author} {\bibfnamefont {A.}~\bibnamefont {Kelly}}, \bibinfo
  {author} {\bibfnamefont {D.~A.}\ \bibnamefont {Cotta}}, \bibinfo {author}
  {\bibfnamefont {B.}~\bibnamefont {Peaudecerf}}, \bibinfo {author}
  {\bibfnamefont {G.~D.}\ \bibnamefont {Bruce}}, \ and\ \bibinfo {author}
  {\bibfnamefont {S.}~\bibnamefont {Kuhr}},\ }\href
  {http://www.nature.com/nphys/journal/v11/n9/full/nphys3403.html} {\bibfield
  {journal} {\bibinfo  {journal} {Nat. Phys.}\ }\textbf {\bibinfo {volume}
  {11}},\ \bibinfo {pages} {738} (\bibinfo {year} {2015})}\BibitemShut
  {NoStop}%
\bibitem [{\citenamefont {Omran}\ \emph {et~al.}(2015)\citenamefont {Omran},
  \citenamefont {Boll}, \citenamefont {Hilker}, \citenamefont {Kleinlein},
  \citenamefont {Salomon}, \citenamefont {Bloch},\ and\ \citenamefont
  {Gross}}]{Omran2015}%
  \BibitemOpen
  \bibfield  {author} {\bibinfo {author} {\bibfnamefont {A.}~\bibnamefont
  {Omran}}, \bibinfo {author} {\bibfnamefont {M.}~\bibnamefont {Boll}},
  \bibinfo {author} {\bibfnamefont {T.~A.}\ \bibnamefont {Hilker}}, \bibinfo
  {author} {\bibfnamefont {K.}~\bibnamefont {Kleinlein}}, \bibinfo {author}
  {\bibfnamefont {G.}~\bibnamefont {Salomon}}, \bibinfo {author} {\bibfnamefont
  {I.}~\bibnamefont {Bloch}}, \ and\ \bibinfo {author} {\bibfnamefont
  {C.}~\bibnamefont {Gross}},\ }\href {\doibase 10.1103/PhysRevLett.115.263001}
  {\bibfield  {journal} {\bibinfo  {journal} {Phys. Rev. Lett.}\ }\textbf
  {\bibinfo {volume} {115}},\ \bibinfo {pages} {263001} (\bibinfo {year}
  {2015})}\BibitemShut {NoStop}%
\bibitem [{\citenamefont {Lester}\ \emph {et~al.}(2015)\citenamefont {Lester},
  \citenamefont {Luick}, \citenamefont {Kaufman}, \citenamefont {Reynolds},\
  and\ \citenamefont {Regal}}]{Lester2015}%
  \BibitemOpen
  \bibfield  {author} {\bibinfo {author} {\bibfnamefont {B.~J.}\ \bibnamefont
  {Lester}}, \bibinfo {author} {\bibfnamefont {N.}~\bibnamefont {Luick}},
  \bibinfo {author} {\bibfnamefont {A.~M.}\ \bibnamefont {Kaufman}}, \bibinfo
  {author} {\bibfnamefont {C.~M.}\ \bibnamefont {Reynolds}}, \ and\ \bibinfo
  {author} {\bibfnamefont {C.~A.}\ \bibnamefont {Regal}},\ }\href {\doibase
  10.1103/PhysRevLett.115.073003} {\bibfield  {journal} {\bibinfo  {journal}
  {Phys. Rev. Lett.}\ }\textbf {\bibinfo {volume} {115}},\ \bibinfo {pages}
  {073003} (\bibinfo {year} {2015})}\BibitemShut {NoStop}%
\bibitem [{\citenamefont {Vassen}\ \emph {et~al.}(2012)\citenamefont {Vassen},
  \citenamefont {Cohen-Tannoudji}, \citenamefont {Leduc}, \citenamefont
  {Boiron}, \citenamefont {Westbrook}, \citenamefont {Truscott}, \citenamefont
  {Baldwin}, \citenamefont {Birkl}, \citenamefont {Cancio},\ and\ \citenamefont
  {Trippenbach}}]{Vassen2012}%
  \BibitemOpen
  \bibfield  {author} {\bibinfo {author} {\bibfnamefont {W.}~\bibnamefont
  {Vassen}}, \bibinfo {author} {\bibfnamefont {C.}~\bibnamefont
  {Cohen-Tannoudji}}, \bibinfo {author} {\bibfnamefont {M.}~\bibnamefont
  {Leduc}}, \bibinfo {author} {\bibfnamefont {D.}~\bibnamefont {Boiron}},
  \bibinfo {author} {\bibfnamefont {C.~I.}\ \bibnamefont {Westbrook}}, \bibinfo
  {author} {\bibfnamefont {A.}~\bibnamefont {Truscott}}, \bibinfo {author}
  {\bibfnamefont {K.}~\bibnamefont {Baldwin}}, \bibinfo {author} {\bibfnamefont
  {G.}~\bibnamefont {Birkl}}, \bibinfo {author} {\bibfnamefont
  {P.}~\bibnamefont {Cancio}}, \ and\ \bibinfo {author} {\bibfnamefont
  {M.}~\bibnamefont {Trippenbach}},\ }\href {\doibase
  10.1103/RevModPhys.84.175} {\bibfield  {journal} {\bibinfo  {journal} {Rev.
  Mod. Physics}\ }\textbf {\bibinfo {volume} {84}},\ \bibinfo {pages} {175}
  (\bibinfo {year} {2012})}\BibitemShut {NoStop}%
\bibitem [{\citenamefont {Perrin}\ \emph {et~al.}(2008)\citenamefont {Perrin},
  \citenamefont {Savage}, \citenamefont {Boiron}, \citenamefont
  {Krachmalnicoff}, \citenamefont {Westbrook},\ and\ \citenamefont
  {Kheruntsyan}}]{Perrin2008}%
  \BibitemOpen
  \bibfield  {author} {\bibinfo {author} {\bibfnamefont {A.}~\bibnamefont
  {Perrin}}, \bibinfo {author} {\bibfnamefont {C.~M.}\ \bibnamefont {Savage}},
  \bibinfo {author} {\bibfnamefont {D.}~\bibnamefont {Boiron}}, \bibinfo
  {author} {\bibfnamefont {V.}~\bibnamefont {Krachmalnicoff}}, \bibinfo
  {author} {\bibfnamefont {C.~I.}\ \bibnamefont {Westbrook}}, \ and\ \bibinfo
  {author} {\bibfnamefont {K.~V.}\ \bibnamefont {Kheruntsyan}},\ }\href
  {http://stacks.iop.org/1367-2630/10/i=4/a=045021} {\bibfield  {journal}
  {\bibinfo  {journal} {New J. Phys.}\ }\textbf {\bibinfo {volume} {10}},\
  \bibinfo {pages} {045021} (\bibinfo {year} {2008})}\BibitemShut {NoStop}%
\bibitem [{\citenamefont {Kheruntsyan}\ \emph {et~al.}(2005)\citenamefont
  {Kheruntsyan}, \citenamefont {Olsen},\ and\ \citenamefont
  {Drummond}}]{Kheruntsyan-EPR-2005}%
  \BibitemOpen
  \bibfield  {author} {\bibinfo {author} {\bibfnamefont {K.~V.}\ \bibnamefont
  {Kheruntsyan}}, \bibinfo {author} {\bibfnamefont {M.~K.}\ \bibnamefont
  {Olsen}}, \ and\ \bibinfo {author} {\bibfnamefont {P.~D.}\ \bibnamefont
  {Drummond}},\ }\href {\doibase 10.1103/PhysRevLett.95.150405} {\bibfield
  {journal} {\bibinfo  {journal} {Phys. Rev. Lett.}\ }\textbf {\bibinfo
  {volume} {95}},\ \bibinfo {pages} {150405} (\bibinfo {year}
  {2005})}\BibitemShut {NoStop}%
\bibitem [{\citenamefont {Lewis-Swan}\ and\ \citenamefont
  {Kheruntsyan}(2015)}]{Lewis-Swan2015}%
  \BibitemOpen
  \bibfield  {author} {\bibinfo {author} {\bibfnamefont {R.~J.}\ \bibnamefont
  {Lewis-Swan}}\ and\ \bibinfo {author} {\bibfnamefont {K.~V.}\ \bibnamefont
  {Kheruntsyan}},\ }\href {\doibase 10.1103/PhysRevA.91.052114} {\bibfield
  {journal} {\bibinfo  {journal} {Phys. Rev. A}\ }\textbf {\bibinfo {volume}
  {91}},\ \bibinfo {pages} {052114} (\bibinfo {year} {2015})}\BibitemShut
  {NoStop}%
\bibitem [{\citenamefont {Lee}\ \emph {et~al.}(2006)\citenamefont {Lee},
  \citenamefont {Chen}, \citenamefont {Liang}, \citenamefont {Li},
  \citenamefont {Voss},\ and\ \citenamefont {Kumar}}]{Lee2006}%
  \BibitemOpen
  \bibfield  {author} {\bibinfo {author} {\bibfnamefont {K.~F.}\ \bibnamefont
  {Lee}}, \bibinfo {author} {\bibfnamefont {J.}~\bibnamefont {Chen}}, \bibinfo
  {author} {\bibfnamefont {C.}~\bibnamefont {Liang}}, \bibinfo {author}
  {\bibfnamefont {X.}~\bibnamefont {Li}}, \bibinfo {author} {\bibfnamefont
  {P.~L.}\ \bibnamefont {Voss}}, \ and\ \bibinfo {author} {\bibfnamefont
  {P.}~\bibnamefont {Kumar}},\ }\href {\doibase 10.1364/OL.31.001905}
  {\bibfield  {journal} {\bibinfo  {journal} {Opt. Lett.}\ }\textbf {\bibinfo
  {volume} {31}},\ \bibinfo {pages} {1905} (\bibinfo {year}
  {2006})}\BibitemShut {NoStop}%
\bibitem [{\citenamefont {He}\ \emph {et~al.}(2012)\citenamefont {He},
  \citenamefont {Satija}, \citenamefont {Clark}, \citenamefont {Rey},\ and\
  \citenamefont {Rigol}}]{He2012}%
  \BibitemOpen
  \bibfield  {author} {\bibinfo {author} {\bibfnamefont {K.}~\bibnamefont
  {He}}, \bibinfo {author} {\bibfnamefont {I.~I.}\ \bibnamefont {Satija}},
  \bibinfo {author} {\bibfnamefont {C.~W.}\ \bibnamefont {Clark}}, \bibinfo
  {author} {\bibfnamefont {A.~M.}\ \bibnamefont {Rey}}, \ and\ \bibinfo
  {author} {\bibfnamefont {M.}~\bibnamefont {Rigol}},\ }\href {\doibase
  10.1103/PhysRevA.85.013617} {\bibfield  {journal} {\bibinfo  {journal} {Phys.
  Rev. A}\ }\textbf {\bibinfo {volume} {85}},\ \bibinfo {pages} {013617}
  (\bibinfo {year} {2012})}\BibitemShut {NoStop}%
\bibitem [{\citenamefont {Rey}\ \emph {et~al.}(2009)\citenamefont {Rey},
  \citenamefont {Sensarma}, \citenamefont {F\"olling}, \citenamefont {Greiner},
  \citenamefont {Demler},\ and\ \citenamefont {Lukin}}]{Rey2009}%
  \BibitemOpen
  \bibfield  {author} {\bibinfo {author} {\bibfnamefont {A.~M.}\ \bibnamefont
  {Rey}}, \bibinfo {author} {\bibfnamefont {R.}~\bibnamefont {Sensarma}},
  \bibinfo {author} {\bibfnamefont {S.}~\bibnamefont {F\"olling}}, \bibinfo
  {author} {\bibfnamefont {M.}~\bibnamefont {Greiner}}, \bibinfo {author}
  {\bibfnamefont {E.}~\bibnamefont {Demler}}, \ and\ \bibinfo {author}
  {\bibfnamefont {M.~D.}\ \bibnamefont {Lukin}},\ }\href
  {http://stacks.iop.org/0295-5075/87/i=6/a=60001} {\bibfield  {journal}
  {\bibinfo  {journal} {EPL (Europhysics Letters)}\ }\textbf {\bibinfo {volume}
  {87}},\ \bibinfo {pages} {60001} (\bibinfo {year} {2009})}\BibitemShut
  {NoStop}%
\bibitem [{\citenamefont {Kunitski}\ \emph {et~al.}(2015)\citenamefont
  {Kunitski}, \citenamefont {Zeller}, \citenamefont {Voigtsberger},
  \citenamefont {Kalinin}, \citenamefont {Schmidt}, \citenamefont
  {Sch{\"o}ffler}, \citenamefont {Czasch}, \citenamefont {Sch{\"o}llkopf},
  \citenamefont {Grisenti}, \citenamefont {Jahnke}, \citenamefont {Blume},\
  and\ \citenamefont {D{\"o}rner}}]{Kunitski2015}%
  \BibitemOpen
  \bibfield  {author} {\bibinfo {author} {\bibfnamefont {M.}~\bibnamefont
  {Kunitski}}, \bibinfo {author} {\bibfnamefont {S.}~\bibnamefont {Zeller}},
  \bibinfo {author} {\bibfnamefont {J.}~\bibnamefont {Voigtsberger}}, \bibinfo
  {author} {\bibfnamefont {A.}~\bibnamefont {Kalinin}}, \bibinfo {author}
  {\bibfnamefont {L.~P.~H.}\ \bibnamefont {Schmidt}}, \bibinfo {author}
  {\bibfnamefont {M.}~\bibnamefont {Sch{\"o}ffler}}, \bibinfo {author}
  {\bibfnamefont {A.}~\bibnamefont {Czasch}}, \bibinfo {author} {\bibfnamefont
  {W.}~\bibnamefont {Sch{\"o}llkopf}}, \bibinfo {author} {\bibfnamefont
  {R.~E.}\ \bibnamefont {Grisenti}}, \bibinfo {author} {\bibfnamefont
  {T.}~\bibnamefont {Jahnke}}, \bibinfo {author} {\bibfnamefont
  {D.}~\bibnamefont {Blume}}, \ and\ \bibinfo {author} {\bibfnamefont
  {R.}~\bibnamefont {D{\"o}rner}},\ }\href {\doibase 10.1126/science.aaa5601}
  {\bibfield  {journal} {\bibinfo  {journal} {Science}\ }\textbf {\bibinfo
  {volume} {348}},\ \bibinfo {pages} {551} (\bibinfo {year}
  {2015})}\BibitemShut {NoStop}%
\bibitem [{\citenamefont {Zi\ifmmode~\acute{n}\else \'{n}\fi{}}\ \emph
  {et~al.}(2005)\citenamefont {Zi\ifmmode~\acute{n}\else \'{n}\fi{}},
  \citenamefont {Chwede\ifmmode~\acute{n}\else \'{n}\fi{}czuk}, \citenamefont
  {Veitia}, \citenamefont {Rza\ifmmode \mbox{\c{}}\else
  \c{}\fi{}\ifmmode~\dot{z}\else \.{z}\fi{}ewski},\ and\ \citenamefont
  {Trippenbach}}]{ZinPRL2005}%
  \BibitemOpen
  \bibfield  {author} {\bibinfo {author} {\bibfnamefont {P.}~\bibnamefont
  {Zi\ifmmode~\acute{n}\else \'{n}\fi{}}}, \bibinfo {author} {\bibfnamefont
  {J.}~\bibnamefont {Chwede\ifmmode~\acute{n}\else \'{n}\fi{}czuk}}, \bibinfo
  {author} {\bibfnamefont {A.}~\bibnamefont {Veitia}}, \bibinfo {author}
  {\bibfnamefont {K.}~\bibnamefont {Rza\ifmmode \mbox{\c{}}\else
  \c{}\fi{}\ifmmode~\dot{z}\else \.{z}\fi{}ewski}}, \ and\ \bibinfo {author}
  {\bibfnamefont {M.}~\bibnamefont {Trippenbach}},\ }\href {\doibase
  10.1103/PhysRevLett.94.200401} {\bibfield  {journal} {\bibinfo  {journal}
  {Phys. Rev. Lett.}\ }\textbf {\bibinfo {volume} {94}},\ \bibinfo {pages}
  {200401} (\bibinfo {year} {2005})}\BibitemShut {NoStop}%
\bibitem [{\citenamefont {Zi\ifmmode~\acute{n}\else \'{n}\fi{}}\ \emph
  {et~al.}(2006)\citenamefont {Zi\ifmmode~\acute{n}\else \'{n}\fi{}},
  \citenamefont {Chwede\ifmmode~\acute{n}\else \'{n}\fi{}czuk},\ and\
  \citenamefont {Trippenbach}}]{ZinPRA2006}%
  \BibitemOpen
  \bibfield  {author} {\bibinfo {author} {\bibfnamefont {P.}~\bibnamefont
  {Zi\ifmmode~\acute{n}\else \'{n}\fi{}}}, \bibinfo {author} {\bibfnamefont
  {J.}~\bibnamefont {Chwede\ifmmode~\acute{n}\else \'{n}\fi{}czuk}}, \ and\
  \bibinfo {author} {\bibfnamefont {M.}~\bibnamefont {Trippenbach}},\ }\href
  {\doibase 10.1103/PhysRevA.73.033602} {\bibfield  {journal} {\bibinfo
  {journal} {Phys. Rev. A}\ }\textbf {\bibinfo {volume} {73}},\ \bibinfo
  {pages} {033602} (\bibinfo {year} {2006})}\BibitemShut {NoStop}%
\bibitem [{\citenamefont {\"Ogren}\ and\ \citenamefont
  {Kheruntsyan}(2009)}]{Oegren2009}%
  \BibitemOpen
  \bibfield  {author} {\bibinfo {author} {\bibfnamefont {M.}~\bibnamefont
  {\"Ogren}}\ and\ \bibinfo {author} {\bibfnamefont {K.~V.}\ \bibnamefont
  {Kheruntsyan}},\ }\href {\doibase 10.1103/PhysRevA.79.021606} {\bibfield
  {journal} {\bibinfo  {journal} {Phys. Rev. A}\ }\textbf {\bibinfo {volume}
  {79}},\ \bibinfo {pages} {021606} (\bibinfo {year} {2009})}\BibitemShut
  {NoStop}%
\bibitem [{\citenamefont {Walls}\ and\ \citenamefont
  {Milburn}(2007)}]{walls2007quantum}%
  \BibitemOpen
  \bibfield  {author} {\bibinfo {author} {\bibfnamefont {D.~F.}\ \bibnamefont
  {Walls}}\ and\ \bibinfo {author} {\bibfnamefont {G.~J.}\ \bibnamefont
  {Milburn}},\ }\href@noop {} {\emph {\bibinfo {title} {Quantum optics}}}\
  (\bibinfo  {publisher} {Springer Science \& Business Media},\ \bibinfo {year}
  {2007})\BibitemShut {NoStop}%
\bibitem [{\citenamefont {Kheruntsyan}\ and\ \citenamefont
  {Drummond}(2002)}]{Kheruntsyan2002}%
  \BibitemOpen
  \bibfield  {author} {\bibinfo {author} {\bibfnamefont {K.~V.}\ \bibnamefont
  {Kheruntsyan}}\ and\ \bibinfo {author} {\bibfnamefont {P.~D.}\ \bibnamefont
  {Drummond}},\ }\href {\doibase 10.1103/PhysRevA.66.031602} {\bibfield
  {journal} {\bibinfo  {journal} {Phys. Rev. A}\ }\textbf {\bibinfo {volume}
  {66}},\ \bibinfo {pages} {031602} (\bibinfo {year} {2002})}\BibitemShut
  {NoStop}%
\bibitem [{\citenamefont {Savage}\ \emph {et~al.}(2006)\citenamefont {Savage},
  \citenamefont {Schwenn},\ and\ \citenamefont
  {Kheruntsyan}}]{Kherunstyan2006Dissociation}%
  \BibitemOpen
  \bibfield  {author} {\bibinfo {author} {\bibfnamefont {C.~M.}\ \bibnamefont
  {Savage}}, \bibinfo {author} {\bibfnamefont {P.~E.}\ \bibnamefont {Schwenn}},
  \ and\ \bibinfo {author} {\bibfnamefont {K.~V.}\ \bibnamefont
  {Kheruntsyan}},\ }\href {\doibase 10.1103/PhysRevA.74.033620} {\bibfield
  {journal} {\bibinfo  {journal} {Phys. Rev. A}\ }\textbf {\bibinfo {volume}
  {74}},\ \bibinfo {pages} {033620} (\bibinfo {year} {2006})}\BibitemShut
  {NoStop}%
\bibitem [{\citenamefont {Lewis-Swan}\ and\ \citenamefont
  {Kheruntsyan}(2014)}]{Lewis-Swan2014}%
  \BibitemOpen
  \bibfield  {author} {\bibinfo {author} {\bibfnamefont {R.~J.}\ \bibnamefont
  {Lewis-Swan}}\ and\ \bibinfo {author} {\bibfnamefont {K.~V.}\ \bibnamefont
  {Kheruntsyan}},\ }\href {http://www.nature.com/articles/ncomms4752}
  {\bibfield  {journal} {\bibinfo  {journal} {Nature Commun.}\ }\textbf
  {\bibinfo {volume} {5}},\ \bibinfo {pages} {3752} (\bibinfo {year}
  {2014})}\BibitemShut {NoStop}%
\bibitem [{\citenamefont {\"Ogren}\ and\ \citenamefont
  {Kheruntsyan}(2010)}]{Oegren2010}%
  \BibitemOpen
  \bibfield  {author} {\bibinfo {author} {\bibfnamefont {M.}~\bibnamefont
  {\"Ogren}}\ and\ \bibinfo {author} {\bibfnamefont {K.~V.}\ \bibnamefont
  {Kheruntsyan}},\ }\href {\doibase 10.1103/PhysRevA.82.013641} {\bibfield
  {journal} {\bibinfo  {journal} {Phys. Rev. A}\ }\textbf {\bibinfo {volume}
  {82}},\ \bibinfo {pages} {013641} (\bibinfo {year} {2010})}\BibitemShut
  {NoStop}%
\bibitem [{\citenamefont {Chwede\ifmmode~\acute{n}\else \'{n}\fi{}czuk}\ \emph
  {et~al.}(2006)\citenamefont {Chwede\ifmmode~\acute{n}\else \'{n}\fi{}czuk},
  \citenamefont {Zi\ifmmode~\acute{n}\else \'{n}\fi{}}, \citenamefont
  {Rza\ifmmode \mbox{\c{}}\else \c{}\fi{}\ifmmode~\dot{z}\else
  \.{z}\fi{}ewski},\ and\ \citenamefont {Trippenbach}}]{ZinPRL2006}%
  \BibitemOpen
  \bibfield  {author} {\bibinfo {author} {\bibfnamefont {J.}~\bibnamefont
  {Chwede\ifmmode~\acute{n}\else \'{n}\fi{}czuk}}, \bibinfo {author}
  {\bibfnamefont {P.}~\bibnamefont {Zi\ifmmode~\acute{n}\else \'{n}\fi{}}},
  \bibinfo {author} {\bibfnamefont {K.}~\bibnamefont {Rza\ifmmode
  \mbox{\c{}}\else \c{}\fi{}\ifmmode~\dot{z}\else \.{z}\fi{}ewski}}, \ and\
  \bibinfo {author} {\bibfnamefont {M.}~\bibnamefont {Trippenbach}},\ }\href
  {\doibase 10.1103/PhysRevLett.97.170404} {\bibfield  {journal} {\bibinfo
  {journal} {Phys. Rev. Lett.}\ }\textbf {\bibinfo {volume} {97}},\ \bibinfo
  {pages} {170404} (\bibinfo {year} {2006})}\BibitemShut {NoStop}%
\bibitem [{\citenamefont {Deuar}\ and\ \citenamefont
  {Drummond}(2007)}]{Deuar2007}%
  \BibitemOpen
  \bibfield  {author} {\bibinfo {author} {\bibfnamefont {P.}~\bibnamefont
  {Deuar}}\ and\ \bibinfo {author} {\bibfnamefont {P.~D.}\ \bibnamefont
  {Drummond}},\ }\href {\doibase 10.1103/PhysRevLett.98.120402} {\bibfield
  {journal} {\bibinfo  {journal} {Phys. Rev. Lett.}\ }\textbf {\bibinfo
  {volume} {98}},\ \bibinfo {pages} {120402} (\bibinfo {year}
  {2007})}\BibitemShut {NoStop}%
\bibitem [{\citenamefont {Krachmalnicoff}\ \emph {et~al.}(2010)\citenamefont
  {Krachmalnicoff}, \citenamefont {Jaskula}, \citenamefont {Bonneau},
  \citenamefont {Leung}, \citenamefont {Partridge}, \citenamefont {Boiron},
  \citenamefont {Westbrook}, \citenamefont {Deuar}, \citenamefont
  {Zi\ifmmode~\acute{n}\else \'{n}\fi{}}, \citenamefont {Trippenbach},\ and\
  \citenamefont {Kheruntsyan}}]{krachmalnicoff2010}%
  \BibitemOpen
  \bibfield  {author} {\bibinfo {author} {\bibfnamefont {V.}~\bibnamefont
  {Krachmalnicoff}}, \bibinfo {author} {\bibfnamefont {J.-C.}\ \bibnamefont
  {Jaskula}}, \bibinfo {author} {\bibfnamefont {M.}~\bibnamefont {Bonneau}},
  \bibinfo {author} {\bibfnamefont {V.}~\bibnamefont {Leung}}, \bibinfo
  {author} {\bibfnamefont {G.~B.}\ \bibnamefont {Partridge}}, \bibinfo {author}
  {\bibfnamefont {D.}~\bibnamefont {Boiron}}, \bibinfo {author} {\bibfnamefont
  {C.~I.}\ \bibnamefont {Westbrook}}, \bibinfo {author} {\bibfnamefont
  {P.}~\bibnamefont {Deuar}}, \bibinfo {author} {\bibfnamefont
  {P.}~\bibnamefont {Zi\ifmmode~\acute{n}\else \'{n}\fi{}}}, \bibinfo {author}
  {\bibfnamefont {M.}~\bibnamefont {Trippenbach}}, \ and\ \bibinfo {author}
  {\bibfnamefont {K.~V.}\ \bibnamefont {Kheruntsyan}},\ }\href {\doibase
  10.1103/PhysRevLett.104.150402} {\bibfield  {journal} {\bibinfo  {journal}
  {Phys. Rev. Lett.}\ }\textbf {\bibinfo {volume} {104}},\ \bibinfo {pages}
  {150402} (\bibinfo {year} {2010})}\BibitemShut {NoStop}%
\bibitem [{\citenamefont {Deuar}\ \emph {et~al.}(2014)\citenamefont {Deuar},
  \citenamefont {Jaskula}, \citenamefont {Bonneau}, \citenamefont
  {Krachmalnicoff}, \citenamefont {Boiron}, \citenamefont {Westbrook},\ and\
  \citenamefont {Kheruntsyan}}]{Deuar2014}%
  \BibitemOpen
  \bibfield  {author} {\bibinfo {author} {\bibfnamefont {P.}~\bibnamefont
  {Deuar}}, \bibinfo {author} {\bibfnamefont {J.-C.}\ \bibnamefont {Jaskula}},
  \bibinfo {author} {\bibfnamefont {M.}~\bibnamefont {Bonneau}}, \bibinfo
  {author} {\bibfnamefont {V.}~\bibnamefont {Krachmalnicoff}}, \bibinfo
  {author} {\bibfnamefont {D.}~\bibnamefont {Boiron}}, \bibinfo {author}
  {\bibfnamefont {C.~I.}\ \bibnamefont {Westbrook}}, \ and\ \bibinfo {author}
  {\bibfnamefont {K.~V.}\ \bibnamefont {Kheruntsyan}},\ }\href {\doibase
  10.1103/PhysRevA.90.033613} {\bibfield  {journal} {\bibinfo  {journal} {Phys.
  Rev. A}\ }\textbf {\bibinfo {volume} {90}},\ \bibinfo {pages} {033613}
  (\bibinfo {year} {2014})}\BibitemShut {NoStop}%
\bibitem [{\citenamefont {Lewis-Swan}(2016)}]{RLS_Thesis}%
  \BibitemOpen
  \bibfield  {author} {\bibinfo {author} {\bibfnamefont {R.~J.}\ \bibnamefont
  {Lewis-Swan}},\ }\href@noop {} {\emph {\bibinfo {title} {Ultracold atoms for
  foundational tests of quantum mechanics}}}\ (\bibinfo  {publisher} {Springer
  International Publishing},\ \bibinfo {year} {2016})\BibitemShut {NoStop}%
\bibitem [{\citenamefont {Bach}\ \emph {et~al.}(2002)\citenamefont {Bach},
  \citenamefont {Trippenbach},\ and\ \citenamefont {Rza\ifmmode
  \mbox{\c{}}\else \c{}\fi{}\ifmmode~\dot{z}\else \.{z}\fi{}ewski}}]{Bach2002}%
  \BibitemOpen
  \bibfield  {author} {\bibinfo {author} {\bibfnamefont {R.}~\bibnamefont
  {Bach}}, \bibinfo {author} {\bibfnamefont {M.}~\bibnamefont {Trippenbach}}, \
  and\ \bibinfo {author} {\bibfnamefont {K.}~\bibnamefont {Rza\ifmmode
  \mbox{\c{}}\else \c{}\fi{}\ifmmode~\dot{z}\else \.{z}\fi{}ewski}},\ }\href
  {\doibase 10.1103/PhysRevA.65.063605} {\bibfield  {journal} {\bibinfo
  {journal} {Phys. Rev. A}\ }\textbf {\bibinfo {volume} {65}},\ \bibinfo
  {pages} {063605} (\bibinfo {year} {2002})}\BibitemShut {NoStop}%
\bibitem [{\citenamefont {Chwede\ifmmode~\acute{n}\else \'{n}\fi{}czuk}\ \emph
  {et~al.}(2008)\citenamefont {Chwede\ifmmode~\acute{n}\else \'{n}\fi{}czuk},
  \citenamefont {Zi\ifmmode~\acute{n}\else \'{n}\fi{}}, \citenamefont
  {Trippenbach}, \citenamefont {Perrin}, \citenamefont {Leung}, \citenamefont
  {Boiron},\ and\ \citenamefont {Westbrook}}]{Chwed2008}%
  \BibitemOpen
  \bibfield  {author} {\bibinfo {author} {\bibfnamefont {J.}~\bibnamefont
  {Chwede\ifmmode~\acute{n}\else \'{n}\fi{}czuk}}, \bibinfo {author}
  {\bibfnamefont {P.}~\bibnamefont {Zi\ifmmode~\acute{n}\else \'{n}\fi{}}},
  \bibinfo {author} {\bibfnamefont {M.}~\bibnamefont {Trippenbach}}, \bibinfo
  {author} {\bibfnamefont {A.}~\bibnamefont {Perrin}}, \bibinfo {author}
  {\bibfnamefont {V.}~\bibnamefont {Leung}}, \bibinfo {author} {\bibfnamefont
  {D.}~\bibnamefont {Boiron}}, \ and\ \bibinfo {author} {\bibfnamefont {C.~I.}\
  \bibnamefont {Westbrook}},\ }\href {\doibase 10.1103/PhysRevA.78.053605}
  {\bibfield  {journal} {\bibinfo  {journal} {Phys. Rev. A}\ }\textbf {\bibinfo
  {volume} {78}},\ \bibinfo {pages} {053605} (\bibinfo {year}
  {2008})}\BibitemShut {NoStop}%
\bibitem [{\citenamefont {Dall}\ and\ \citenamefont
  {Truscott}(2007)}]{Dall2007}%
  \BibitemOpen
  \bibfield  {author} {\bibinfo {author} {\bibfnamefont {R.}~\bibnamefont
  {Dall}}\ and\ \bibinfo {author} {\bibfnamefont {A.}~\bibnamefont
  {Truscott}},\ }\href {\doibase
  http://dx.doi.org/10.1016/j.optcom.2006.09.031} {\bibfield  {journal}
  {\bibinfo  {journal} {Opt. Commun.}\ }\textbf {\bibinfo {volume} {270}},\
  \bibinfo {pages} {255 } (\bibinfo {year} {2007})}\BibitemShut {NoStop}%
\bibitem [{\citenamefont {Manning}\ \emph {et~al.}(2015)\citenamefont
  {Manning}, \citenamefont {Khakimov}, \citenamefont {Dall},\ and\
  \citenamefont {Truscott}}]{Manning2015_Wheeler}%
  \BibitemOpen
  \bibfield  {author} {\bibinfo {author} {\bibfnamefont {A.~G.}\ \bibnamefont
  {Manning}}, \bibinfo {author} {\bibfnamefont {R.~I.}\ \bibnamefont
  {Khakimov}}, \bibinfo {author} {\bibfnamefont {R.~G.}\ \bibnamefont {Dall}},
  \ and\ \bibinfo {author} {\bibfnamefont {A.~G.}\ \bibnamefont {Truscott}},\
  }\href {\doibase 10.1038/nphys3343} {\bibfield  {journal} {\bibinfo
  {journal} {Nat. Phys.}\ }\textbf {\bibinfo {volume} {11}},\ \bibinfo {pages}
  {539} (\bibinfo {year} {2015})}\BibitemShut {NoStop}%
\bibitem [{\citenamefont {Kapitza}\ and\ \citenamefont
  {Dirac}(1933)}]{Kapitza1933}%
  \BibitemOpen
  \bibfield  {author} {\bibinfo {author} {\bibfnamefont {P.~L.}\ \bibnamefont
  {Kapitza}}\ and\ \bibinfo {author} {\bibfnamefont {P.~A.~M.}\ \bibnamefont
  {Dirac}},\ }\href {\doibase 10.1017/S0305004100011105} {\bibfield  {journal}
  {\bibinfo  {journal} {Math. Proc. Cambridge Philos. Soc.}\ }\textbf {\bibinfo
  {volume} {29}},\ \bibinfo {pages} {297} (\bibinfo {year} {1933})}\BibitemShut
  {NoStop}%
\bibitem [{\citenamefont {Gould}\ \emph {et~al.}(1986)\citenamefont {Gould},
  \citenamefont {Ruff},\ and\ \citenamefont {Pritchard}}]{Gould1986}%
  \BibitemOpen
  \bibfield  {author} {\bibinfo {author} {\bibfnamefont {P.~L.}\ \bibnamefont
  {Gould}}, \bibinfo {author} {\bibfnamefont {G.~A.}\ \bibnamefont {Ruff}}, \
  and\ \bibinfo {author} {\bibfnamefont {D.~E.}\ \bibnamefont {Pritchard}},\
  }\href {\doibase 10.1103/PhysRevLett.56.827} {\bibfield  {journal} {\bibinfo
  {journal} {Phys. Rev. Lett.}\ }\textbf {\bibinfo {volume} {56}},\ \bibinfo
  {pages} {827} (\bibinfo {year} {1986})}\BibitemShut {NoStop}%
\bibitem [{\citenamefont {Ovchinnikov}\ \emph {et~al.}(1999)\citenamefont
  {Ovchinnikov}, \citenamefont {M\"uller}, \citenamefont {Doery}, \citenamefont
  {Vredenbregt}, \citenamefont {Helmerson}, \citenamefont {Rolston},\ and\
  \citenamefont {Phillips}}]{Ovchinnikov1999}%
  \BibitemOpen
  \bibfield  {author} {\bibinfo {author} {\bibfnamefont {Y.~B.}\ \bibnamefont
  {Ovchinnikov}}, \bibinfo {author} {\bibfnamefont {J.~H.}\ \bibnamefont
  {M\"uller}}, \bibinfo {author} {\bibfnamefont {M.~R.}\ \bibnamefont {Doery}},
  \bibinfo {author} {\bibfnamefont {E.~J.~D.}\ \bibnamefont {Vredenbregt}},
  \bibinfo {author} {\bibfnamefont {K.}~\bibnamefont {Helmerson}}, \bibinfo
  {author} {\bibfnamefont {S.~L.}\ \bibnamefont {Rolston}}, \ and\ \bibinfo
  {author} {\bibfnamefont {W.~D.}\ \bibnamefont {Phillips}},\ }\href {\doibase
  10.1103/PhysRevLett.83.284} {\bibfield  {journal} {\bibinfo  {journal} {Phys.
  Rev. Lett.}\ }\textbf {\bibinfo {volume} {83}},\ \bibinfo {pages} {284}
  (\bibinfo {year} {1999})}\BibitemShut {NoStop}%
\bibitem [{\citenamefont {Deuar}\ \emph {et~al.}(2013)\citenamefont {Deuar},
  \citenamefont {Wasak}, \citenamefont {Zi\ifmmode~\acute{n}\else \'{n}\fi{}},
  \citenamefont {Chwede\ifmmode~\acute{n}\else \'{n}\fi{}czuk},\ and\
  \citenamefont {Trippenbach}}]{Deuar2013}%
  \BibitemOpen
  \bibfield  {author} {\bibinfo {author} {\bibfnamefont {P.}~\bibnamefont
  {Deuar}}, \bibinfo {author} {\bibfnamefont {T.}~\bibnamefont {Wasak}},
  \bibinfo {author} {\bibfnamefont {P.}~\bibnamefont {Zi\ifmmode~\acute{n}\else
  \'{n}\fi{}}}, \bibinfo {author} {\bibfnamefont {J.}~\bibnamefont
  {Chwede\ifmmode~\acute{n}\else \'{n}\fi{}czuk}}, \ and\ \bibinfo {author}
  {\bibfnamefont {M.}~\bibnamefont {Trippenbach}},\ }\href {\doibase
  10.1103/PhysRevA.88.013617} {\bibfield  {journal} {\bibinfo  {journal} {Phys.
  Rev. A}\ }\textbf {\bibinfo {volume} {88}},\ \bibinfo {pages} {013617}
  (\bibinfo {year} {2013})}\BibitemShut {NoStop}%
\bibitem [{\citenamefont {Zambelli}\ \emph {et~al.}(2000)\citenamefont
  {Zambelli}, \citenamefont {Pitaevskii}, \citenamefont {Stamper-Kurn},\ and\
  \citenamefont {Stringari}}]{Zambelli2000}%
  \BibitemOpen
  \bibfield  {author} {\bibinfo {author} {\bibfnamefont {F.}~\bibnamefont
  {Zambelli}}, \bibinfo {author} {\bibfnamefont {L.}~\bibnamefont
  {Pitaevskii}}, \bibinfo {author} {\bibfnamefont {D.~M.}\ \bibnamefont
  {Stamper-Kurn}}, \ and\ \bibinfo {author} {\bibfnamefont {S.}~\bibnamefont
  {Stringari}},\ }\href {\doibase 10.1103/PhysRevA.61.063608} {\bibfield
  {journal} {\bibinfo  {journal} {Phys. Rev. A}\ }\textbf {\bibinfo {volume}
  {61}},\ \bibinfo {pages} {063608} (\bibinfo {year} {2000})}\BibitemShut
  {NoStop}%
\bibitem [{\citenamefont {Hodgman}\ \emph {et~al.}(2011)\citenamefont
  {Hodgman}, \citenamefont {Dall}, \citenamefont {Manning}, \citenamefont
  {Baldwin},\ and\ \citenamefont {Truscott}}]{Hodgman2011}%
  \BibitemOpen
  \bibfield  {author} {\bibinfo {author} {\bibfnamefont {S.~S.}\ \bibnamefont
  {Hodgman}}, \bibinfo {author} {\bibfnamefont {R.~G.}\ \bibnamefont {Dall}},
  \bibinfo {author} {\bibfnamefont {A.~G.}\ \bibnamefont {Manning}}, \bibinfo
  {author} {\bibfnamefont {K.~G.~H.}\ \bibnamefont {Baldwin}}, \ and\ \bibinfo
  {author} {\bibfnamefont {A.~G.}\ \bibnamefont {Truscott}},\ }\href {\doibase
  10.1126/science.1198481} {\bibfield  {journal} {\bibinfo  {journal}
  {Science}\ }\textbf {\bibinfo {volume} {331}},\ \bibinfo {pages} {1046}
  (\bibinfo {year} {2011})}\BibitemShut {NoStop}%
\bibitem [{\citenamefont {Campbell}\ \emph {et~al.}(2015)\citenamefont
  {Campbell}, \citenamefont {Gangardt},\ and\ \citenamefont
  {Kheruntsyan}}]{Campbell:2015}%
  \BibitemOpen
  \bibfield  {author} {\bibinfo {author} {\bibfnamefont {A.~S.}\ \bibnamefont
  {Campbell}}, \bibinfo {author} {\bibfnamefont {D.~M.}\ \bibnamefont
  {Gangardt}}, \ and\ \bibinfo {author} {\bibfnamefont {K.~V.}\ \bibnamefont
  {Kheruntsyan}},\ }\href {\doibase 10.1103/PhysRevLett.114.125302} {\bibfield
  {journal} {\bibinfo  {journal} {Phys. Rev. Lett.}\ }\textbf {\bibinfo
  {volume} {114}},\ \bibinfo {pages} {125302} (\bibinfo {year}
  {2015})}\BibitemShut {NoStop}%
\bibitem [{\citenamefont {Hong}\ \emph {et~al.}(1987)\citenamefont {Hong},
  \citenamefont {Ou},\ and\ \citenamefont {Mandel}}]{HOM}%
  \BibitemOpen
  \bibfield  {author} {\bibinfo {author} {\bibfnamefont {C.~K.}\ \bibnamefont
  {Hong}}, \bibinfo {author} {\bibfnamefont {Z.~Y.}\ \bibnamefont {Ou}}, \ and\
  \bibinfo {author} {\bibfnamefont {L.}~\bibnamefont {Mandel}},\ }\href
  {\doibase 10.1103/PhysRevLett.59.2044} {\bibfield  {journal} {\bibinfo
  {journal} {Phys. Rev. Lett.}\ }\textbf {\bibinfo {volume} {59}},\ \bibinfo
  {pages} {2044} (\bibinfo {year} {1987})}\BibitemShut {NoStop}%
\end{thebibliography}%

\clearpage


\renewcommand{\thefigure}{S\arabic{figure}}
\renewcommand{\theequation}{S\arabic{equation}}
\renewcommand{\thepage}{S\arabic{page}}

\setcounter{figure}{0} 
\setcounter{equation}{0} 
\setcounter{page}{1} 


\section*{Supplemental Material}


\section*{1. Theoretical background and connection to experimental quantities}

In many-body theory, multi-particle correlations are characterised by the $N$-th order ($N$-point) correlation function which, in the normalized and normally-ordered form, can be expressed in terms of the creation and annihilation quantum field operators, $\hat{\Psi}^{\dagger}(\mathbf{x})$ and $ \hat{\Psi}(\mathbf{x})$, as
\begin{multline}
g^{(N)}(\mathbf{x}_1,\mathbf{x}_2,...,\mathbf{x}_N)=\frac{\langle \,: \!\hat{n}(\mathbf{x}_1) \hat{n}(\mathbf{x}_2) ... \,\hat{n}(\mathbf{x}_N) \!:\,\rangle}{\langle \hat{n}(\mathbf{x}_1) \rangle \langle \hat{n}(\mathbf{x}_2) \rangle ... \, \langle \hat{n}(\mathbf{x}_N) \rangle} \\
 =\frac{\langle \hat{\Psi}^{\dagger}(\mathbf{x}_1) \hat{\Psi}^{\dagger}(\mathbf{x}_2) ... \hat{\Psi}^{\dagger}(\mathbf{x}_N) \hat{\Psi}(\mathbf{x}_N) ... \hat{\Psi}(\mathbf{x}_2) \hat{\Psi}(\mathbf{x}_1) \rangle}{\langle \hat{n}(\mathbf{x}_1) \rangle \langle \hat{n}(\mathbf{x}_2) \rangle ... \langle \hat{n}(\mathbf{x}_N) \rangle}. \label{corr}
\end{multline}
Here, we have considered equal-time density [$\hat{n}(\mathbf{x}_j)\!\equiv\! \hat{\Psi}^{\dagger}(\mathbf{x}_j) \hat{\Psi}(\mathbf{x}_j)$, $j\!=\!1,2,...,N$] correlations in position space for simplicity, but the definition and the context can be extended to, e.g., correlations in the reciprocal momentum space. The physical meaning of the above correlation function is that it represents the joint probability of detecting $N$ particles in their respective positions $\{\mathbf{x}_1, \mathbf{x}_2,..., \mathbf{x}_N\}$, normalized to the product of single-particle detection probabilities.

Operationally, sufficient knowledge of higher-order correlation functions is equivalent to solving the quantum many-body problem. For example, consider the case of a thermal state, which is an example of a broad class of Gaussian states.  Here, all higher-order correlation functions factorize (as per Wick's theorem) into a sum of terms involving only products of normal densities $\langle \hat{\Psi}^{\dagger}(\mathbf{x}_i) \hat{\Psi}(\mathbf{x}_j) \rangle$ ($i,j=1,2,...,N$).  Hence their knowledge alone is sufficient to predict all these higher-order correlations. For the same-point correlation function $g^{(N)}(\mathbf{x},\mathbf{x},...,\mathbf{x})$, this property leads to the simple result that $g^{(N)}(\mathbf{x},\mathbf{x},...,\mathbf{x})=N!$, which for $N=2$ implies the Hanbury Brown--Twiss bunching effect \cite{Jeltes2007}. More generally, Wick's factorisation for Gaussian states requires inclusion of terms also containing products of anomalous densities, $\langle \hat{\Psi}(\mathbf{x}_i) \hat{\Psi}(\mathbf{x}_j) \rangle$, which do  not have to be zero in general (unlike for thermal states). The anomalous density (also known as the anomalous Green's function in quantum field theory) involves a product of two annihilation operators and characterises the amplitude of nonlocal pair correlations present in the system.

Momentum-space correlations studied in this work can be defined analogously to Eq.~(\ref{corr}) as
\begin{eqnarray}
 g^{(N)}(\mathbf{k}_1,\mathbf{k}_2,...,\mathbf{k}_N) & = & \frac{\langle \,: \!\hat{n}_{\mathbf{k}_1} \hat{n}_{\mathbf{k}_2} ... \,\hat{n}_{\mathbf{k}_N} \!:\,\rangle}{\langle \hat{n}_{\mathbf{k}_1} \rangle \langle \hat{n}_{\mathbf{k}_2} \rangle ... \, \langle \hat{n}_{\mathbf{k}_N} \rangle}   \notag \\
& = & \frac{\langle \hat{a}^{\dagger}_{\mathbf{k}_1}\hat{a}^{\dagger}_{\mathbf{k}_2}...\hat{a}^{\dagger}_{\mathbf{k}_N} \hat{a}_{\mathbf{k}_N} ... \hat{a}_{\mathbf{k}_2}\hat{a}_{\mathbf{k}_1} \rangle}{\langle \hat{n}_{\mathbf{k}_1} \rangle \langle \hat{n}_{\mathbf{k}_2} \rangle ... \, \langle \hat{n}_{\mathbf{k}_N} \rangle}  , \label{eq:gN_momspace}
\end{eqnarray}
with $\hat{a}^{\dagger}_{\mathbf{k}}$ and $\hat{a}_{\mathbf{k}}$ being the momentum mode creation and annihilation operators in plane-wave basis, satisfying equal-time bosonic  commutation relation 
$\left[\hat{a}_{\mathbf{k}},\hat{a}^{\dagger}_{\mathbf{k}'}\right] = \delta_{\mathbf{k},\mathbf{k}'}$, and $n_{\mathbf{k}}\equiv \langle \hat{n}_{\mathbf{k}}\rangle=\langle \hat{a}^{\dagger}_{\mathbf{k}} \hat{a}_{\mathbf{k}} \rangle$ giving the distribution of halo mode occupancies. We have omitted the explicit time-dependence of the operators for notational simplicity.

The scattering problem under investigation can be described within the undepleted and constant (in time) source BEC approximation, valid when the total number of scattered atoms constitutes only a small fraction ($<5\%$) of the initial number of atoms in the source condensates and when the mean kinetic energy of the colliding atoms is much larger than the mean interaction energy per atom \cite{ZinPRL2005,ZinPRA2006,Oegren2009}. Furthermore, given that the initial trapping frequencies in all three dimensions are rather weak in our experiments and that the scattering occurs predominantly from the condensate central (high-density) region where the real-space density is nearly constant, we can approximate the source condensate as a uniform system, in which case the scattered atoms can be described by the following pairwise coupled Heisenberg equations of motion \cite{Oegren2009}:
\begin{eqnarray}
\frac{d\hat{a}_{\mathbf{k}}(t)}{dt} & = & -i\Delta _{k}\hat{a}_{\mathbf{k}}(t)-i\chi\hat{a}^{\dagger }_{\mathbf{-k}}(t),  \label{HeisenbergEquation-2a} \\
\frac{d\hat{a}^{\dag}_{\mathbf{-k}}(t)}{dt}& = &i\Delta _{k}\hat{a}_{\mathbf{-k}}(t)+i\chi\hat{a}_{\mathbf{k}}(t).  \label{HeisenbergEquation-2b}
\end{eqnarray}
Here, $\chi=U\bar{\rho}/\hbar$ is the effective coupling, $U = 4\pi\hbar^2a_s/m$ is the $s$-wave interaction strength characterised by the scattering length $a_s$, and $\bar{\rho}$ is the density of the source condensate prior to its splitting into two colliding (counterpropagating) halves. In addition, $\Delta _{k}\equiv \hbar (k^{2}-k_{r}^{2})/\left( 2m\right) $ is the effective detuning from the scattering resonance, with $k^{2}=|%
\mathbf{k}|^{2}$ and $k_r$ the collision momentum, i.e., the momentum kick imparted onto each colliding condensate in the center-of-mass frame.

\begin{table*}[ht!]
 \centering
 \caption{Complete set of 4th order correlation functions in terms of normal and anomalous lattice mode occupancies, $n_{\mathbf{k}_i}$ and $m_{\mathbf{k}_i}$. All momenta $\mathbf{k}_i$ ($i=1,2,3,4$) are drawn from the halo peak centered radially at $|\mathbf{k}_i|=k_r$, and for notational simplicity we have omitted the momentum index (i.e. $n_{\mathbf{k}_i} \equiv n$ and $m_{\mathbf{k}_i} \equiv m$). The unnormalised correlation function is denoted via $G^{(4)}(\mathbf{k}_1,\mathbf{k}_2,\mathbf{k}_3,\mathbf{k}_4)$ and refers to the numerator in Eq. (\ref{eq:gN_momspace}).} 
 \label{tab:g3g4}
 \begin{tabular}{| c | c | c | c | }

  \hline
  Case & $G^{(4)}(\mathbf{k}_1,\mathbf{k}_2,\mathbf{k}_3,\mathbf{k}_4)$ & $g^{(4)}(\mathbf{k}_1,\mathbf{k}_2,\mathbf{k}_3,\mathbf{k}_4)$ & Decays to case: \\ \hline
  (1) \vspace{0cm} \includegraphics[valign=m,scale=\scaleVal,trim=0 0 0 -5]{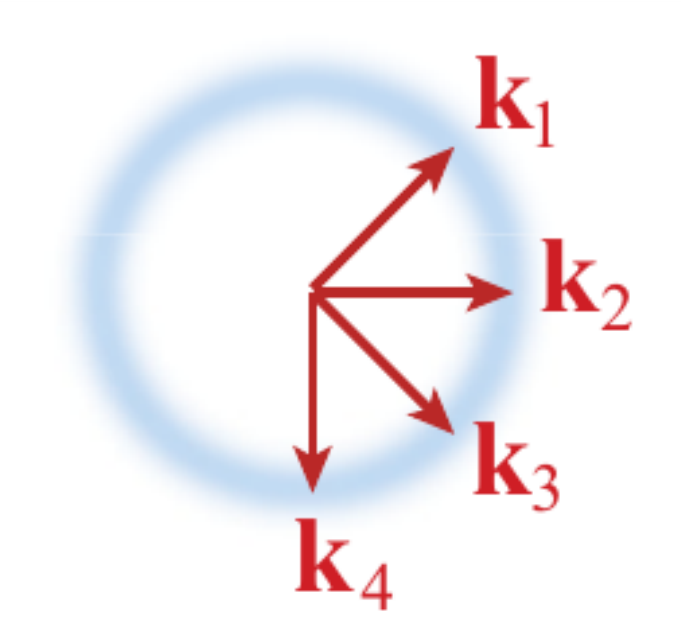} & $n^4$ & $1$ & N/A \\ \hline
  (2) \vspace{0cm} \includegraphics[valign=m,scale=\scaleVal,trim=0 0 0 -5]{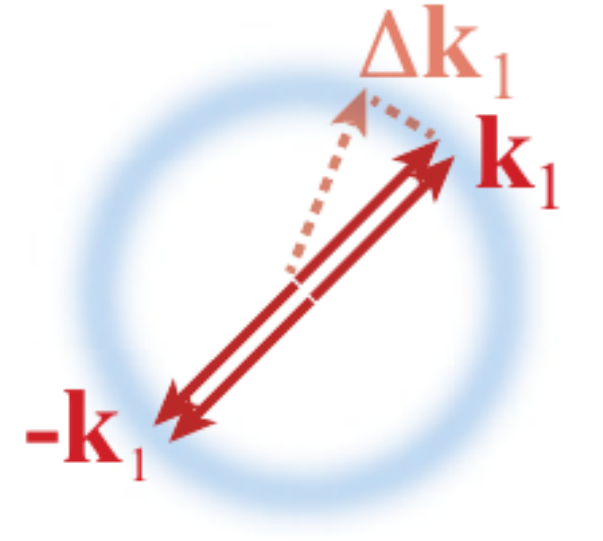} & $4|m|^4 + 16n^2|m|^2 + 4n^4$ & $24 + 24/n + 4/n^2$ & (3) \\ \hline
  (3) \vspace{0cm} \includegraphics[valign=m,scale=\scaleVal,trim=0 0 0 -5]{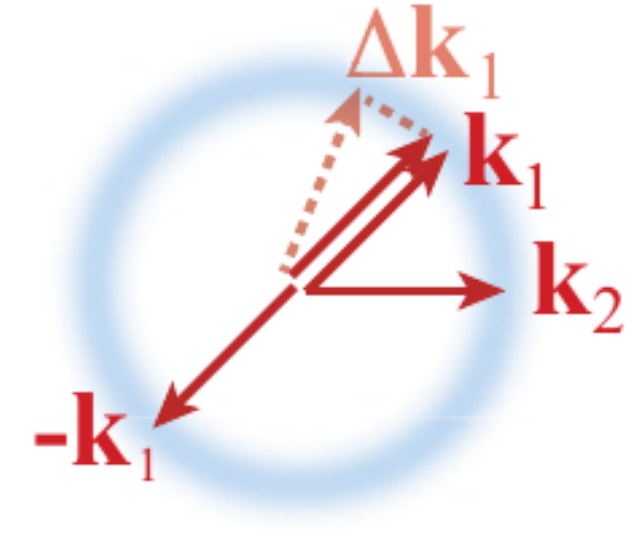} & $2n^2(2|m|^2 + n^2)$ & $6 + 4/n$ & (6) \\ \hline
  (4) \vspace{0cm} \includegraphics[valign=m,scale=\scaleVal,trim=0 0 0 -5]{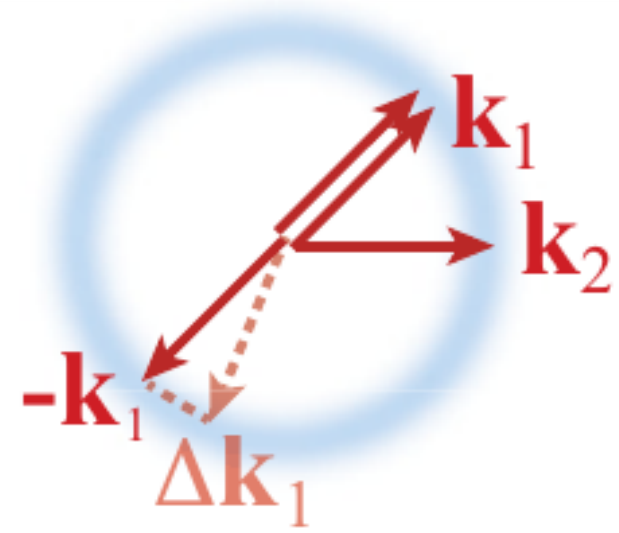} & $2n^2(2|m|^2 + n^2)$ & $6 + 4/n$ & (5) \\ \hline
  (5) \vspace{0cm} \includegraphics[valign=m,scale=\scaleVal,trim=0 0 0 -5]{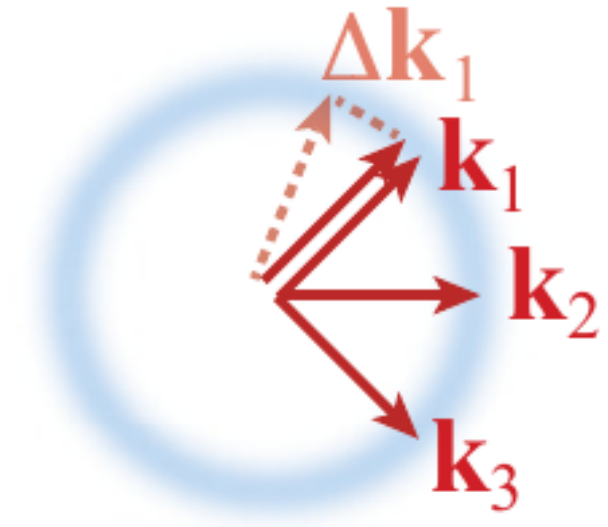} & $2n^4$ & $2$ & (1) \\ \hline
  (6) \vspace{0cm} \includegraphics[valign=m,scale=\scaleVal,trim=0 0 0 -5]{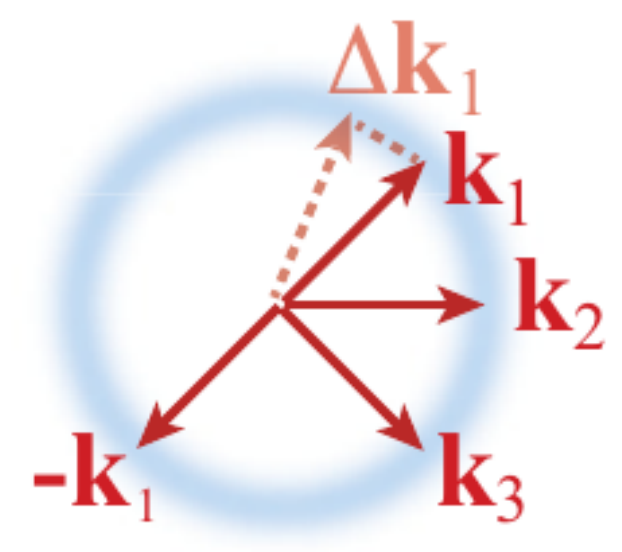} & $n^2(n^2 + |m|^2)$ & $2 + 1/n$ & (1) \\ \hline
  (7) \vspace{0cm} \includegraphics[valign=m,scale=\scaleVal,trim=0 0 0 -5]{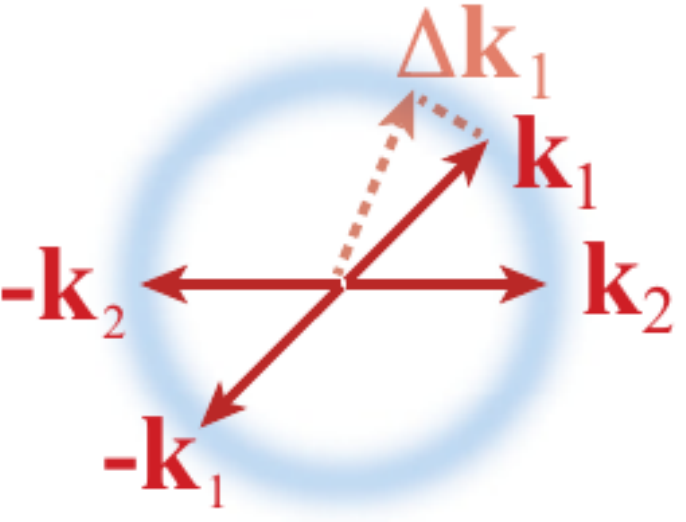} & $(n^2 + |m|^2)^2$ & $4 + 4/n + 1/n^2$ & (6) \\ \hline
  (8) \vspace{0cm} \includegraphics[valign=m,scale=\scaleVal,trim=0 0 0 -5]{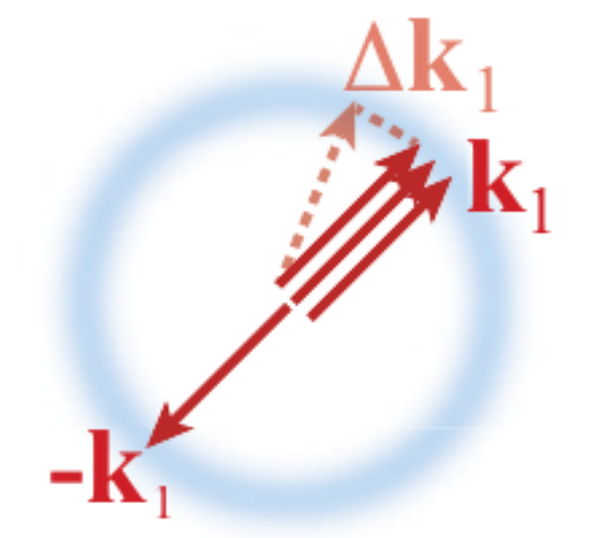} & $6n^2(n^2 + 3|m|^2)$ & $24 + 18/n$ & (3) \\ \hline
  (9) \vspace{0cm} \includegraphics[valign=m,scale=\scaleVal,trim=0 0 0 -5]{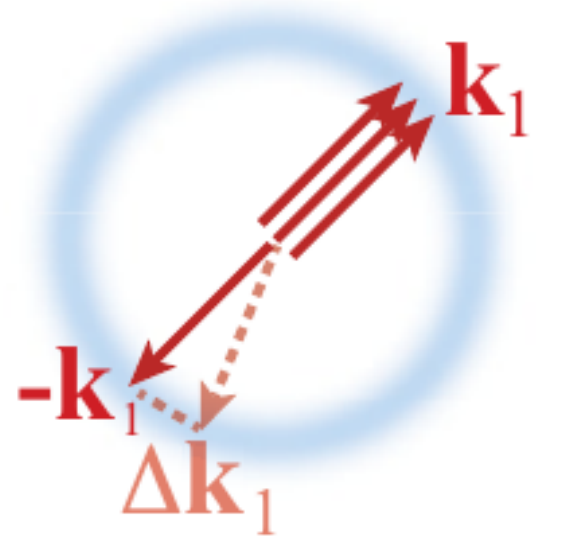} & $6n^2(n^2 + 3|m|^2)$ & $24 + 18/n$ & (10) \\ \hline
  (10) \vspace{0cm} \includegraphics[valign=m,scale=\scaleVal,trim=0 0 0 -5]{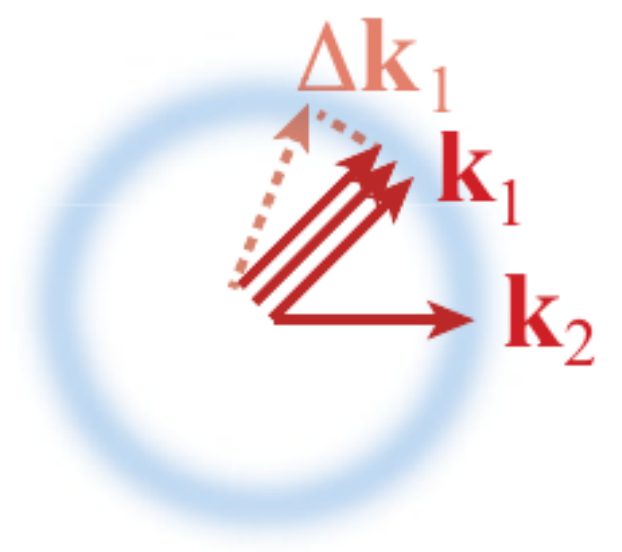} & $6n^4$ & $6$ & (5) \\ \hline
  (11) \vspace{0cm} \includegraphics[valign=m,scale=\scaleVal,trim=0 0 0 -5]{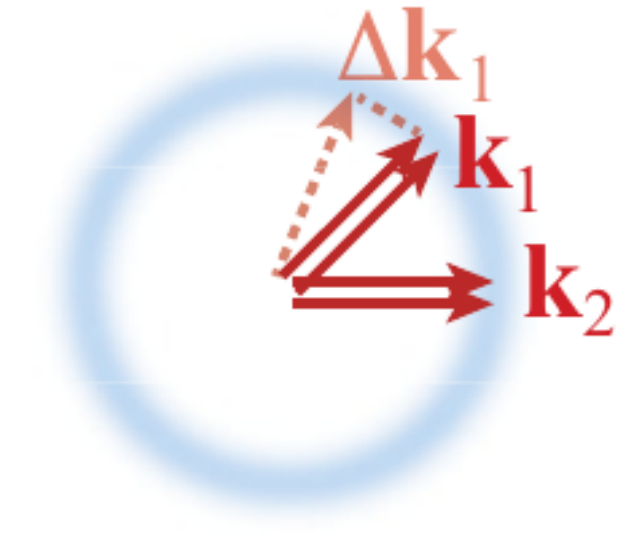} & $4n^4$ & $4$ & (5) \\ \hline
  (12) \vspace{0cm} \includegraphics[valign=m,scale=\scaleVal,trim=0 0 0 -5]{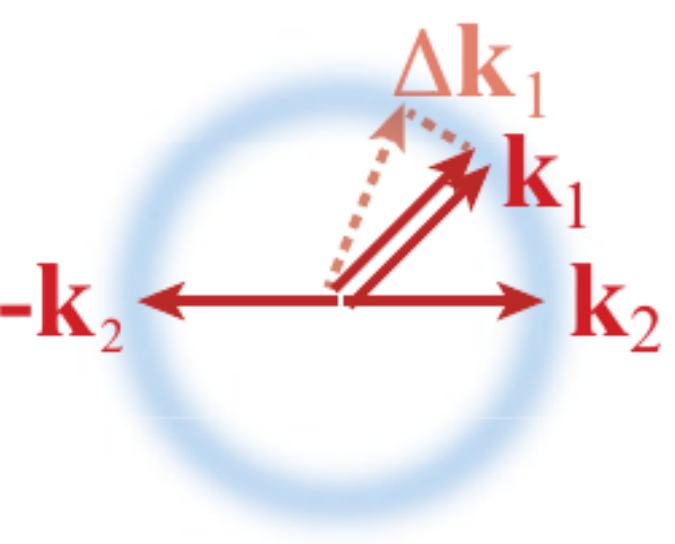} & $2n^2(n^2 + |m|^2)$ & $6 + 4/n$ & (6) \\ \hline
  (13) \vspace{0cm} \includegraphics[valign=m,scale=\scaleVal,trim=0 0 0 -5]{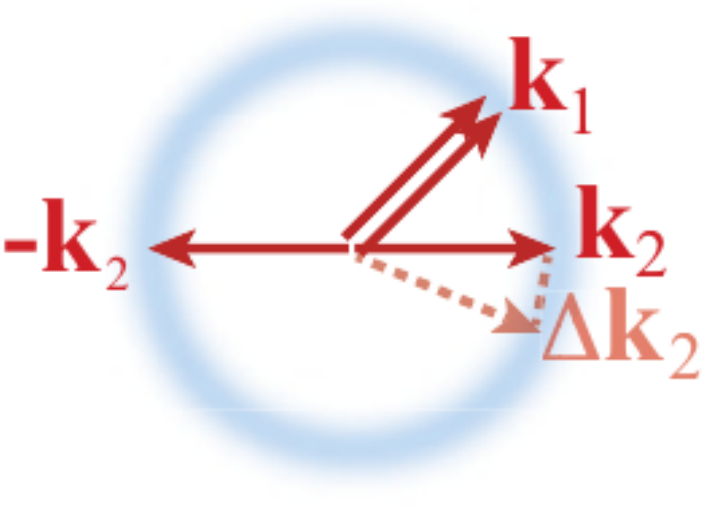} & $2n^2(n^2 + |m|^2)$ & $6 + 4/n$ & (5) \\ \hline
  (14) \vspace{0cm} \includegraphics[valign=m,scale=\scaleVal,trim=0 0 0 -5]{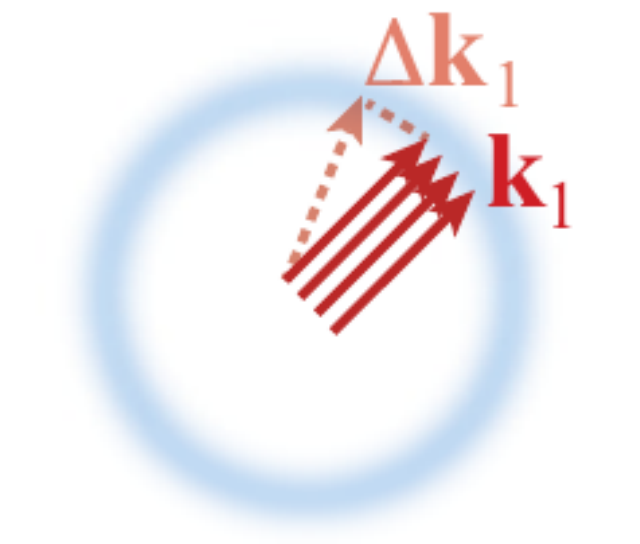} & $24n^4$ & $24$ & (10) \\ \hline
 \end{tabular}
\end{table*}

The explicit solutions to these equations, which are analogous to the equations of motion for producing a two-mode squeezed vacuum state in quantum optics by means of parametric downconversion \cite{walls2007quantum}, are \cite{Kheruntsyan2002,Kherunstyan2006Dissociation,Lewis-Swan2014}:
\begin{eqnarray}
 \hat{a}_{\mathbf{k}}(t) & = & \alpha_{\mathbf{k}}(t)\hat{a}_{\mathbf{k}}(0) + \beta_{\mathbf{k}}(t)\hat{a}^{\dagger}_{-\mathbf{k}}(0) , \label{eq:ak_scat} \\
 \hat{a}^{\dagger}_{-\mathbf{k}}(t) & = & \beta^*_{\mathbf{k}}(t)\hat{a}_{\mathbf{k}}(0) + \alpha^*_{\mathbf{k}}(t)\hat{a}^{\dagger}_{-\mathbf{k}}(0), \label{eq:amk_scat}
\end{eqnarray}
where the coefficients $\alpha_{\mathbf{k}}(t)$ and $\beta_{\mathbf{k}}(t)$ are given by 
\begin{eqnarray}
 \alpha_{\mathbf{k}}(t) & = & \left[\mathrm{cosh}\left(\sqrt{\chi^2 - \Delta^2_k}\,t\right) \right. \notag \\ 
 & & \left. + \frac{i\Delta_k}{\sqrt{\chi^2 - \Delta^2_k}}\mathrm{sinh}\left(\sqrt{\chi^2 - \Delta^2_k}\,t\right)\right] e^{i\frac{\hbar |\mathbf{k}_r|^2}{2m}t} , \\
 \beta_{\mathbf{k}}(t) & = & \frac{-i\chi}{\sqrt{\chi^2 - \Delta^2_k}}\mathrm{sinh}\left(\sqrt{\chi^2 - \Delta^2_k}\,t\right) e^{i\frac{\hbar |\mathbf{k}_r|^2}{2m}t} ,
\end{eqnarray}
and satisfy $|\beta_{\mathbf{k}}|^2-|\alpha_{\mathbf{k}}|^2=1$.

From these solutions, and for vacuum initial conditions for all scattering modes, one finds that the only nonzero second-order moments of creation and annihilation operators in this model are the normal and anomalous mode occupancies given, respectively, by:
\begin{eqnarray}
n_{\mathbf{k}}(t) & = &\langle \hat{a}^{\dag}_{\mathbf{k}}(t)  \hat{a}_{\mathbf{k}}(t) \rangle = |\beta_{\mathbf{k}}(t)|^2 , \label{nk}\\
m_{\mathbf{k}}(t) & = &\langle \hat{a}_{\mathbf{k}}(t)  \hat{a}_{-\mathbf{k}}(t) \rangle = \alpha_{\mathbf{k}}(t) \beta_{\mathbf{k}}(t). \label{mk}
\end{eqnarray}

An important relationship that follows immediately from these solutions is that the amplitude of the anomalous occupancy $|m_{\mathbf{k}}|$ for any mode is related to the normal occupancy $n_{\mathbf{k}}$ by
\begin{equation}
|m_{\mathbf{k}}|^2=n_{\mathbf{k}}\left(n_{\mathbf{k}}+1\right),
\label{relationship-discrete}
\end{equation}
where  $n_{\mathbf{k}}(t)=\frac{\chi^2}{\chi^2 - \Delta^2_k}\sinh^2\left(\sqrt{\chi^2 - \Delta^2_k}\,t\right)$.  This follows from the bosonic commutation relation and the conservation of particle number difference in equal but opposite momentum modes in the $s$-wave pair-production process. Note that for any resonant (or peak) halo mode $\mathbf{k}_0$, for which $\Delta_k\!=\!0$ and therefore $|\mathbf{k}_0|\!=\!k_r$, the mode occupation acquires the well-known form for the two-mode squeezed vacuum state, $n_{\mathbf{k_0}}(t)\!=\!\sinh^2\left(\chi\,t\right)$ \cite{walls2007quantum,Kheruntsyan2002,Kherunstyan2006Dissociation}.

The linearity of Eqs.~(\ref{HeisenbergEquation-2a})--(\ref{HeisenbergEquation-2b}) ensures that the evolution will result in a Gaussian many-body state and hence Wick's factorisation scheme for all higher-order correlation functions will apply. This allows one to express any $N$-point correlation as a sum of terms containing all possible distinct products of the nonzero second-order moments. Taking the second-order correlation function, i.e., for $N\!=\!2$ in Eq.~(\ref{eq:gN_momspace}), between two collinear (CL) atoms with the same momenta as an example, this means that it can be evaluated as 
\begin{eqnarray}
g_{CL}^{(2)}(\mathbf{k}) & \equiv & g^{(2)}(\mathbf{k},\mathbf{k})=\frac{\langle :\hat{n}^{\dagger}_{\mathbf{k}}\hat{n}^{\dagger}_{\mathbf{k}} :\rangle}{\langle \hat{n}_{\mathbf{k}} \rangle \langle \hat{n}_{\mathbf{k}} \rangle }=\frac{\langle \hat{a}^{\dagger}_{\mathbf{k}}\hat{a}^{\dagger}_{\mathbf{k}} \hat{a}_{\mathbf{k}}\hat{a}_{\mathbf{k}} \rangle}{\langle \hat{n}_{\mathbf{k}} \rangle \langle \hat{n}_{\mathbf{k}} \rangle } \notag \\
& = &\frac{2n_{\mathbf{k}}^2}{n_{\mathbf{k}}^2}=2.  \label{eq:g2_bb_analytic_a}
\end{eqnarray}%
Similarly, the back-to-back (BB) second-order correlation function between two atoms with equal but opposite momenta will be given by
\begin{eqnarray}
g^{(2)}_{BB}(\mathbf{k}) & \equiv & g^{(2)}(\mathbf{k},-\mathbf{k}) = \frac{\langle :\hat{n}^{\dagger}_{\mathbf{k}}\hat{n}^{\dagger}_{-\mathbf{k}} :\rangle}{\langle \hat{n}_{\mathbf{k}} \rangle \langle \hat{n}_{-\mathbf{k}} \rangle }=\frac{\langle \hat{a}^{\dagger}_{\mathbf{k}}\hat{a}^{\dagger}_{-\mathbf{k}} \hat{a}_{-\mathbf{k}}\hat{a}_{\mathbf{k}} \rangle}{\langle \hat{n}_{\mathbf{k}} \rangle \langle \hat{n}_{-\mathbf{k}} \rangle } \notag \\
& = & \frac{n_{\mathbf{k}}^2 + |m_{\mathbf{k}}|^2}{n_{\mathbf{k}}^2} = 2 + \frac{1}{n_{\mathbf{k}}}. \label{eq:g2_bb_analytic_b}
\end{eqnarray}

The third- and any higher-order peak correlations and their expressions in terms of the mode occupancy $n_{\mathbf{k}}$ can be evaluated similarly;
the complete set of third-order correlations is listed in Fig.~2 of the main text, while we list fourteen nontrivial fourth-order correlations in Table~\ref{tab:g3g4}.

Quantitatively, the best agreement between the results based on the uniform source-condensate model and those that can be obtained numerically for nonuniform condensates is achieved when the size of the quantisation box in the uniform model is matched with the characteristic size of the source \cite{Kherunstyan2006Dissociation,Oegren2009,Oegren2010}. In this case, the  spacing between the plane-wave modes $\Delta k= 2\pi/L$ (which we assume is for a cubic box of side $L$, for simplicity) takes the role of the coherence or correlation length beyond which any higher-order correlation function containing pairs of nearly equal or nearly 
opposite momenta decays (stepwise) from the peak value to the respective uncorrelated value. For nonuniform systems, on the other hand, the physical correlation length may span over several plane-wave modes depending on the quantization volume adopted for the problem. In this case, the results of the uniform model, while being applicable to peak amplitudes of correlation functions for nonuniform systems, cannot address the quantitative details of the decay of these correlations with distance. Such details can be addressed numerically using stochastic approaches in phase-space \cite{ZinPRL2006,Deuar2007,krachmalnicoff2010,Deuar2014,Lewis-Swan2014,RLS_Thesis} or analytic approaches based on perturbation theory \cite{Bach2002,ZinPRA2006,Chwed2008,Oegren2009,RLS_Thesis}, which are, however, beyond the scope of this work.

Thus, in order to use the simple analytic results of this section for comparison with experimentally measured quantities (see below), we will restrict ourselves to quantitative results that refer to the peak amplitudes of correlation functions. Furthermore, we note that in order to improve the counting statistics the experimental data is averaged over a certain spherical shell (excluding the regions occupied by the source BECs), which is justified due to the spherical symmetry of the scattering halo and the fact that the halo mode size in our experiments is comparable with the radial thickness of the shell. This means that the average halo-mode occupancy ($n$), used in the analysis of experimental results, is identified with 
\begin{equation}
n=\frac{1}{M}\sum_{\mathbf{k} \in V_S} \langle \hat{n}_{\mathbf{k}} \rangle =  \frac{1}{M}\sum_{\mathbf{k} \in V_S} n_{\mathbf{k}}   \simeq n_{\mathbf{k}_0},
\label{average-n}
\end{equation}
in our analytic theory; similarly, the absolute value of the average anomalous occupancy $|m|$ refers to
\begin{equation}
|m|=\frac{1}{M}\sum_{\mathbf{k} \in V_S}  |m_{\mathbf{k}}|   \simeq |m_{\mathbf{k}_0}|.
\label{average-m}
\end{equation}
Here, $V_S$ is the averaging volume chosen to be a spherical shell centered (radially) at the halo peak $|\mathbf{k}_0|=k_r$, whereas $M$ is the number of modes contained in $V_S$ (see the next section for details).

\section*{2. Experimental details} 

The experiment begins with a He$^*$ BEC trapped in a bi-planar quadrupole Ioffe configuration magnetic trap~\cite{Dall2007}.  The trap has harmonic frequencies of $\{\omega_x, \omega_y, \omega_z\}/2\pi\!\simeq\!\{15,25,25\}$\,Hz and a bias magnetic field $B_0\!=\!1.31(1)$~G along the $\hat{\mathbf{x}}$-axis.  Note that the near uniformity of the trapping frequencies justifies the angular integration performed to convert $g_{BB}^{(3)}(\Delta \mathbf{k}_1,\Delta \mathbf{k}_2)$ to $\bar{g}_{BB}^{(3)}(\Delta k_1,\Delta k_2)$ described in the main text.  Our detector consists of a pair of $80$\,mm diameter multichannel plates and a delay line, located $\sim\!850$\,mm below the trap center ($416$\,ms fall time), which has a quantum efficiency of $\approx$10$\%$.  

\subsection*{2.1 Halo generation} 

Similarly to our previous work~\cite{Manning2015_Wheeler} we employ the same laser beams for both Raman and Bragg pulses, only changing the relative frequency detuning of the waveforms, which is set by the bias~$B_0$ and geometrical angle between the beams~(90$^\circ$). 

\subsubsection*{Bragg diffraction} 

We use scattering off a grating formed by the same two laser beams in either the Raman, Bragg and Kapitza-Dirac regimes to generate the halos, similar to our previous work \cite{Manning2015_Wheeler,Khakimov2016}.  The laser beams are blue detuned by $\approx$ 2~GHz from the $2^3$S$_1\rightarrow 2^3$P$_0$ transition.  For all experiments we start with a Raman pulse to transfer $\approx 95 \%$ of the BEC atoms in the $\mathbf{k}=0$ momentum mode of the $m_J=+1$ sublevel into the magnetically insensitive $m_J=0$ sublevel.  These untrapped atoms are then diffracted using a second pulse into a number of diffraction orders, depending on the desired mode occupancy of the final $s$-wave halo as follows. For the halos with the five highest mode occupancies, the second diffraction pulse uses Bragg scattering to transfer between $\sim\!10\%$ and $\sim\!~50\%$ of the atoms back into the $\mathbf{k}=0$ momentum mode.  The two momentum modes then collide, producing a sphere of scattered atom pairs 
through $s$-wave collisions \cite{Perrin2007}.  The sphere has a radius in momentum space approximately equal to 
$k_r\simeq k_0/\sqrt{2}$ (note that there is a small deviation due to the halo not being quite spherical \cite{krachmalnicoff2010,Deuar2014} and the angle between the diffraction beams not being exactly 90$^{\circ}$).  This allows us to produce halos with average mode occupancies ranging from $n=0.104(8)$ to $n=0.44(2)$.  Here $n=N/M$, where $N$ is the average number of atoms per halo and $M$ is the average number of modes for that halo (see below). The halos for this subset of the data are produced using five different population transfer fractions, with $7$,$305$ shots for each transfer fraction.

\begin{figure*}[tbp]

	\centering

	\includegraphics[width=\textwidth, keepaspectratio=true]{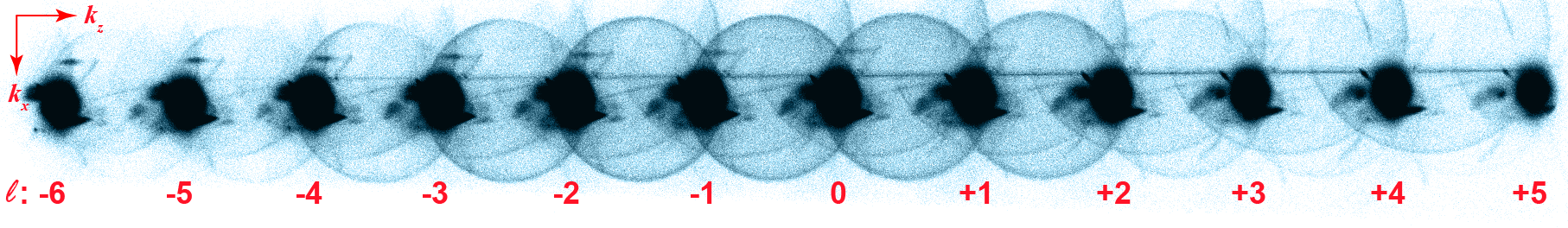}
	\caption{ {\bf Kapitza-Dirac scattering halos.}  Experimental data showing the reconstructed momentum space density for all 11 halos produced by Kapitza-Dirac scattering.  The black regions show the diffraction orders of the BECs, which saturate the detector, with diffraction indices $l$ labeled in red.  Scattered halo atoms are shown in blue. }
	\label{KD_halos}
\end{figure*}

\subsubsection*{Kapitza-Dirac diffraction} 

To generate halos with smaller mode occupancies, we use higher-order Kapitza-Dirac scattering~\cite{Kapitza1933,Gould1986,Ovchinnikov1999} to produce multiple BECs that each act as distinct sources of $s$-wave scattering halos in an individual experimental run \cite{Khakimov2016}.  This yields 12 separate diffraction orders $l=-6...+5$, with each adjacent pair then colliding to produce 11 different halos (see Fig. \ref{KD_halos}).  Each halo still has a radius of $k_r$ in momentum space, centred in between the diffracted condensates. As the diffraction orders are populated with varying numbers of atoms, according to the Kapitza-Dirac effect, a consequence is that the average mode occupancy in the resulting halos ($n^{l,l+1}$) ranges from $n^{0,+1}=0.077(10)$ down to $n^{+5,+6}=0.0017(17)$.  By creating and measuring $11$ different halo mode occupancies in a single experimental run, this increases the data acquisition rate by over an order of magnitude, which is crucial to achieve good signal to noise for halos with such a low number of counts per shot. The Kapitza-Dirac data comes from $54$,$473$ separate individual experimental runs.  This means the total number of different individual halos presented in this manuscript is $635$,$728$.

For both methods of generating $s$-wave scattering halos, it is convenient to operate in momentum space co-ordinates with the origin at the centre of each halo as follows.  Knowing the time-of-flight to the detector $T_{f}$, we convert measured atom spatial coordinates~$\mathbf{r}$ in the detector plane to momenta~$\mathbf{k}$ via  $\mathbf{r}= T_{f}\mathbf{v}$, using $\mathbf{v} = \hbar \mathbf{k}/m_{He}$, where $m_{He}$ is the mass of a ${}^4$He*~atom.

\subsubsection*{Stimulated versus spontaneous scattering}

The combination of the two halo generation methods produces a dataset of halos where the mode occupancy spans more than two orders of magnitude.  
An important parameter that characterises whether the scattering is in the spontaneous or stimulated regime is the ratio of two characteristic timescales 
$\gamma=t_{col}/t_{int}$, the collision duration $t_{col}$ and the interaction timescale $t_{int}$. The characteristic collision duration itself is determined by the smallest of two timescales, $t_{col}=\min\{t_{exp},t_{sep}\}$: the characteristic timescale for expansion of the colliding BECs $t_{exp}\simeq 1/\omega_{y,z}$ (which in our experiment is determined by the largest trapping frequency) and the characteristic timescale for geometric separation of the colliding condensates $t_{sep}\simeq 2R_{TF}/3v_r$, with $R_{TF}$ being the Thomas-Fermi radius of the BEC in the collision direction ($\sim70$\,$\mu$m) and $v_{r}=\hbar k_{r}/m_{He}$ the collision velocity. The numerical factor of $2/3$ in $t_{sep}$ is introduced to account for the fact that, as the colliding condensates separate in space, the density-dependent scattering rate \cite{ZinPRL2005,Perrin2008} becomes significantly smaller well before the condensate overlap region vanishes completely. With these definitions and for our experimental parameters, we estimate that 
$t_{exp}\sim 8$\,ms, whereas $t_{sep}\sim 0.7$\,ms, and therefore the characteristic collision duration is $t_{col}\sim 0.7$\,ms. 

The characteristic interaction timescale, on the other hand, is given by $t_{int}\simeq 1/\chi_0=\hbar/U\rho_0$ and defines the intrinsic pair production rate should the scattering proceed at a constant effective coupling strength $\chi_0$, in the simple undepleted and uniform condensate model of Section 1, Eqs. (\ref{eq:ak_scat})-(\ref{eq:amk_scat}). Within this model, where the halo peak mode occupancy grows as $n_{\mathbf{k}}(t)=\sinh^2(\chi_0t)$, the spontaneous scattering regime corresponds to $\chi_0t\ll1$ where $n_{\mathbf{k}}(t)\simeq (\chi_0t)^2$ grows nearly quadratically with time, whereas the stimulated regime corresponds to $\chi_0t>1$ where the growth of $n_{\mathbf{k}}(t)$ becomes near exponential. The predictions of this model will remain approximately valid as long as the condensate density in the overlap region is not significantly diminished due to their expansion in free space or their geometric spatial separation. Accordingly, if the stimulated regime is never reached due to $t_{int}$ remaining smaller than the actual collision duration $t_{col}$, the pair production process will remain in the spontaneous regime. Estimating $t_{int}$ for our experiment, with $\chi_0=U\bar{\rho}/\hbar$, $U=4\pi\hbar^2 a_s/m_{He}$, $a_s=7.51$\,nm, and taking $\bar{\rho}$ equal to the average BEC density within the typical parabolic Thomas-Fermi profile, we obtain (for our highest density samples with peak density of $\rho(0)\simeq 1.7\times 10^{18}$\,m$^{-3}$) $t_{int}\sim 1$\,ms. This gives $t_{col}/t_{int}\simeq 0.7$. 

Thus, for the halos with maximum mode occupancy, the largest value for our experiment is $\gamma \simeq 0.7$, meaning it should always be in the spontaneous regime of pair production.  Increasing $\gamma > 1$ would cause the system to enter the stimulated regime, where Bose-enhancement leads to a few modes dominating the scattering and a range of other effects, such as the formation of phase grains, become apparent \cite{Deuar2013}. However, as long as the condensates are not significantly depleted, the Wick's factorisation scheme described in Section 1 still holds and we would expect to observe similar correlation properties.

\subsection*{2.2 Halo mode occupancy}

The mode occupancy in the halo is given by $n=N/M$, where $N$ is the total number of atoms in the halo and $M$ is the number of modes in the halo.  Following Ref.~\cite{Perrin2008}, we take $M=V_S/V_M$, where 
\begin{equation}\label{eq:ShellVolume}
	V_S\simeq 4\pi\sqrt{2\pi}k_r^2w 
\end{equation}
is the momentum-space volume of the scattering shell (assuming a Gaussian profile radially), with $w$ the rms width of the shell and $k_r$ the shell radius.  The mode volume, $V_M$, is given by
\begin{equation}\label{eq:ModeVolume}
	V_M\simeq (2\pi)^{3/2}(\sigma_{k})^3,
\end{equation}
where $\sigma_{k}$ is the rms width of the momentum distribution of the source BEC, assuming that it can be approximated by an isotropic Gaussian distribution.  For a simple Thomas-Fermi parabolic density profile of the source BEC (which our BECs can be well approximated by), the momentum distribution is given by a Bessel function \cite{Zambelli2000,Oegren2009}; the bulk of the Bessel function can be fitted by a Gaussian, which is then used to define the rms width $\sigma_k$ in Eq.~(\ref{eq:ModeVolume}). 

Previous analytic and numerical calculations of the second-order back-to-back correlation function have demonstrated the relation $\sigma_{k} \approx \sigma_{BB}/1.1$ \cite{Oegren2009}, which motivates us to use the experimentally measured values of $\sigma_{BB}$ (see the discussion in the following section) to extract $\sigma_{k}$. Note that although $\sigma_{k}$ will be slightly different along each dimension corresponding to the magnetic trap axes, the spherically averaged correlation function measured experimentally in this paper yields only a single, spherically averaged correlation length $\sigma_{BB}$.  The assumption of an isotropic correlation length, i.e. that it is the same along all directions, is an approximation. However, there is only a small asymmetry in our trap frequencies, which in our previous work \cite{Khakimov2016} led to a measured variation of $<15\%$ in the correlation lengths between different axes and thus the approximation should be valid.

From the average Gaussian halo thickness (measured across all halos) of $w_{av} \approx 0.031k_r$ and the fitted back-to-back correlation length 
we find $M$ ranging from 2,900(200)--32,000(21,000).

\begin{figure*}[tbp]

	\centering

	\includegraphics[width=\textwidth, keepaspectratio=true]{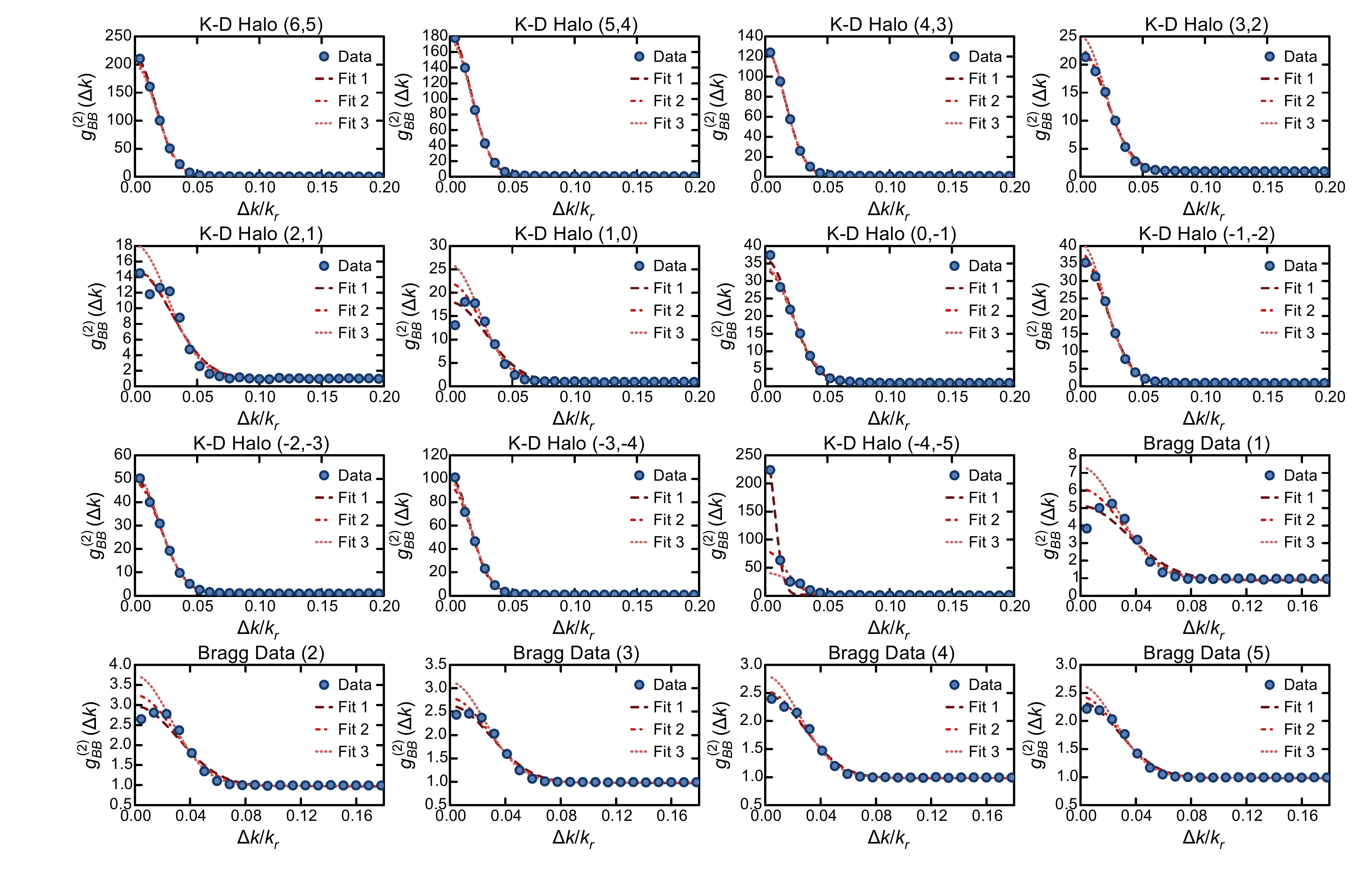}
	\caption{ {\bf Individual plots of back-to-back two-atom correlation functions.}  Full plots and fits of $\bar{g}_{BB}^{(2)}(\Delta k)$ for all 11 halos produced by Kapitza-Dirac (labeled K-D in the figure) and the 5 halos produced by Bragg scattering, along with fit functions.  See text for details of the fits.  Fit 1 is Eq.~(\ref{eq:g2bb_fit}) fitted to all data, fit 2 excludes the smallest $\Delta k$ value and fit 3 excludes the smallest 2 $\Delta k$ values.}
	\label{g2_fits}
\end{figure*}

\subsection*{2.3 Correlation functions}

\subsubsection*{Back-to-back correlations}

From the aggregate (over many repeated runs of the experiment) detected atoms within each experimental run, we can reconstruct the normalised two-atom back-to-back correlation function,
\begin{equation}\label{eq:g2bb}
	g_{BB}^{(2)}(\Delta \mathbf{k}) = 
	\frac{\sum_{\mathbf{k} \in V} \langle : \hat{n}_{\mathbf{k}} \hat{n}_{\mathbf{-k}+\Delta \mathbf{k}} : \rangle}
	     {\sum_{\mathbf{k} \in V}  \langle \hat{n}_{\mathbf{k}}\rangle \langle \hat{n}_{-\mathbf{k}+\Delta\mathbf{k}}\rangle }, 
\end{equation}
between atom pairs with nearly equal but opposite momenta, $\mathbf{k}$ and $ -\mathbf{k} + \Delta \mathbf{k}$.
Due to the spherically symmetric nature of the scattering halo, the summation over momenta $\mathbf{k}$ lying within a certain spherical shell of volume $V$ corresponds to first averaging the unnormalised correlation function over the bulk of the halo (to improve the overall statistics) and then normalising the result with respect to the uncorrelated counterpart of the same quantity. Thus, the normalisation and averaging in Eq. (\ref{eq:g2bb}) ensures that $g_{BB}^{(2)}(\Delta \mathbf{k})=1$ for uncorrelated momenta, while any dependence on $\mathbf{k}$ within $V$ is lost due to the averaging.

The integration volume $V$ in Eq.~(\ref{eq:g2bb}), as in all correlation functions described in this paper, is performed over a thin shell in momentum space encompassing the halo but excluding the condensates and any background atoms which lie outside the halo.  Radially, this volume includes all atoms with $0.55 k_r\!<\!|\mathbf{k}|\!<\!1.28 k_r$, which means all atoms in the halo (gaussian width $\approx \!0.03 k_r$) will be counted.  The volume around the condensates is windowed out by taking $-0.44k_r\!<k_z<\!0.44k_r$, which is somewhat conservative but avoids potential saturation problems that we observe on the delay-line detector when the flux becomes too high.  Note that although the experimental integration volume $V$ is somewhat thicker radially than $V_S$, very few correlated pairs are measured outside $V_S$.  Therefore the correlation amplitudes are dominated by atoms in $V_S$ and thus the use of Eq.~(\ref{average-n}) in theoretical predictions will yield a reasonable approximation for comparison with the experimental results.

Furthermore, for consistency across all correlation functions and to allow easy visualisation of higher-order functions, we additionally integrate over the angles of the displacement vector $\Delta \mathbf{k}$ in spherical coordinates, both for the top and bottom lines of Eq.~(\ref{eq:g2bb}). We denote the resulting correlation function as $\bar{g}_{BB}^{(2)}(\Delta k)$, which we plot (see Fig. \ref{g2_fits} for examples) as a function of the scalar distance between atoms $\Delta k=|\Delta \mathbf{k}|$.

Similarly, the three-atom back-to-back correlation function is given by
\begin{multline}
\label{eq:g3bbbbcl}
	g_{BB}^{(3)}(\Delta \mathbf{k}_1,\Delta \mathbf{k}_2) = 
	\frac{\sum_{\mathbf{k}_3 \in V}  \langle : \hat{n}_{\mathbf{k}_3} \hat{n}_{\mathbf{-k}_3+\Delta \mathbf{k}_1} \hat{n}_{\mathbf{-k}_3+\Delta \mathbf{k}_2} : \rangle}
	     {\sum_{\mathbf{k}_3 \in V}  \langle \hat{n}_{\mathbf{k}_3}\rangle \langle \hat{n}_{-\mathbf{k}_3+\Delta\mathbf{k}_1}\rangle \langle \hat{n}_{-\mathbf{k}_3+\Delta\mathbf{k}_2}\rangle} .
\end{multline} 
Again, to allow clear visualisation of the correlation function, as well as improve the signal to noise, it is more practical to plot the correlations as a function of the scalar distances $\Delta k_1=|\Delta \mathbf{k}_1|$ and $\Delta k_2=|\Delta \mathbf{k}_2|$, which involves spherically integrating over all angles for both the top and bottom terms of Eq.~(\ref{eq:g3bbbbcl}).  We denote the corresponding correlation function $\bar{g}_{BB}^{(3)}(\Delta k_1,\Delta k_2)$, and this function is plotted in Fig.
2 (g) of the main text.

\begin{figure}[bp]
\includegraphics[width=0.47\textwidth, keepaspectratio=true]{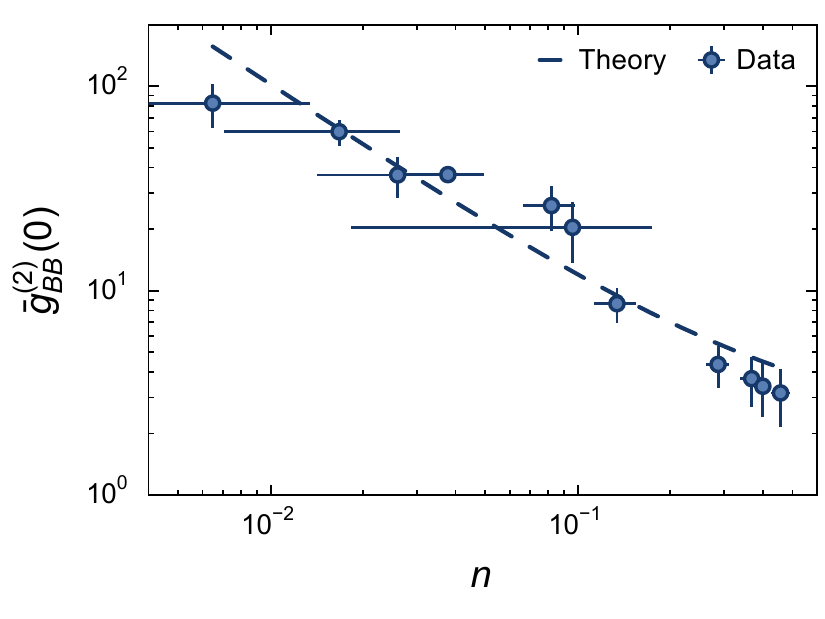}
	\caption{ {\bf  Two-atom back-to-back correlation amplitudes extracted from three-atom correlation functions.}  The two-atom correlation amplitudes $\bar{g}^{(2)}_{BB}(0)$ as extracted from the asymptotes of the full three-atom correlation functions $\bar{g}^{(3)}_{BB}(\Delta k_1,\Delta k_2)$ (blue lines in Fig.
2), plotted along against the average mode occupancy $n$. The dotted line shows the analytic theory curve.  This plot can be directly compared to Fig.~
3 (a). }
	\label{g2_from_g3}
\end{figure}

As a consistency check, we can use the inter-dependency of the different correlation functions to reconstruct $\bar{g}_{BB}^{(2)}(\Delta k)$ from the asymptotes of the full three-atom correlation function $\bar{g}_{BB}^{(3)}(\Delta k_1,\Delta k_2)$.  This is highlighted by the blue lines $\bar{g}_{BB}^{(3)}(\Delta k_1,\Delta k_2\gg \sigma_{BB})$ and $\bar{g}_{BB}^{(3)}(\Delta k_1\gg \sigma_{BB},\Delta k_2)$ in  Fig.
2 (g).  We average the two lines and fit a Gaussian to extract $\bar{g}_{BB}^{(2)}(0)$. These extracted values are shown in Fig.~\ref{g2_from_g3} along with the theory curve Eq.~(1).  The agreement between experiment and theory is comparable to the values of $\bar{g}_{BB}^{(2)}(0)$ measured directly 
[Fig.~
3 (a)].

\subsubsection*{Collinear correlations}  

Similarly to Eq.~(\ref{eq:g2bb}), the collinear two-particle correlation function is defined as
\begin{equation}\label{eq:g2cl}
	g_{CL}^{(2)}(\Delta \mathbf{k}) = 	\frac{\sum_{\mathbf{k} \in V} \langle : \hat{n}_{\mathbf{k}} \hat{n}_{\mathbf{k}+\Delta \mathbf{k}} : \rangle}
	     {\sum_{\mathbf{k} \in V} \langle \hat{n}_{\mathbf{k}}\rangle \langle \hat{n}_{\mathbf{k}+\Delta\mathbf{k}}\rangle }, 
\end{equation}
with the same spherical integration as described above performed to transform from $g_{CL}^{(2)}(\Delta \mathbf{k})$   to  $\bar{g}_{CL}^{(2)}(\Delta k)$, where $\Delta k=|\Delta \mathbf{k}|$.  Due to the low number of counts for some values of $n$ the fits have a large uncertainty and in fact for two halos the statistics are so poor that no meaningful fit is possible.  However, theoretically there is no expected trend with $n$, as in the limit of small detector resolution and small correlation function bins we expect $\bar{g}_{CL}^{(2)}(0)\simeq 2$ for all $n$ \cite{Perrin2008,Oegren2009}.

The correlation function between three collinear atoms is then analogously defined as
\begin{multline}\label{eq:g3cl}
	g_{CL}^{(3)}(\Delta \mathbf{k}_1,\Delta \mathbf{k}_2) \!=\! 
	\frac{\sum_{\mathbf{k}_3 \in V} \langle : \hat{n}_{\mathbf{k}_3}\hat{n}_{\mathbf{k}_3+\Delta \mathbf{k}_1} \hat{n}_{\mathbf{k}_3+\Delta \mathbf{k}_2} : \rangle}
	     {\sum_{\mathbf{k}_3 \in V}  \langle \hat{n}_{\mathbf{k}_3}\rangle \langle \hat{n}_{\mathbf{k}_3+\Delta\mathbf{k}_1}\rangle \langle \hat{n}_{\mathbf{k}_3+\Delta\mathbf{k}_2}\rangle},
\end{multline} 
where now $\mathbf{k}_1= \mathbf{k}_3 + \Delta \mathbf{k}_1$ and $\mathbf{k}_2= \mathbf{k}_3 + \Delta \mathbf{k}_2$.  As with Eq.~(\ref{eq:g3bbbbcl}), the actual correlation function we plot is integrated over all angles and denoted $\bar{g}_{CL}^{(3)}(\Delta k_1,\Delta k_2)$, which is shown in Fig.~
2 \,(h) of the main text. This surface plot is equivalent to previous measurements of $\bar{g}^{(3)}(\Delta k_1,\Delta k_2)$ for cold thermal Bose gases \cite{Hodgman2011,Dall2013_idealnbody}, although note that in the present experiment our BEC source would be much colder and without the collision would not display such correlations \cite{Hodgman2011}.

\subsubsection*{Higher-order correlations}

The poor statistics associated with three-atom collinear correlations in the regimes investigated in this manuscript mean that not enough values of $n$ yield $\bar{g}_{CL}^{(3)}(\Delta k_1,\Delta k_2)$ with sufficient signal to extract the dependence of $\bar{g}_{CL}^{(3)}(0,0)$ on $n$.  However, theoretically we expect $\bar{g}_{CL}^{(3)}(0,0)=6$ for all values of $n$. Even for three-atom back-to-back correlations $\bar{g}_{BB}^{(3)}(\Delta k_1,\Delta k_2)$, reliable correlation functions can only be extracted for 11 different values of $n$, hence only 11 points are shown in Fig.~
3\,(c) of the main text.

\begin{figure}[bp]	\includegraphics[width=0.47\textwidth, keepaspectratio=true]{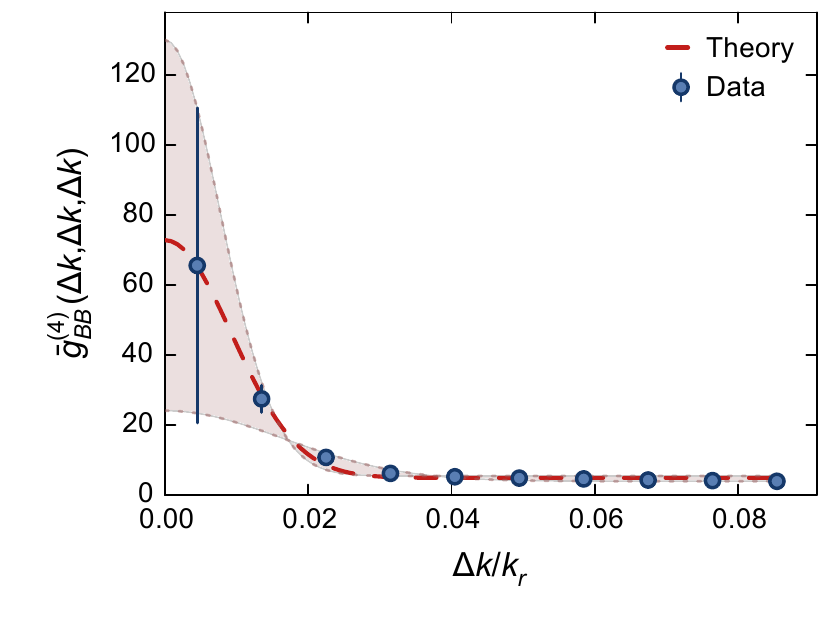}
	\caption{ {\bf  Four-atom back-to-back correlation function.}  Full four-atom correlation function $\bar{g}^{(4)}_{BB}(\Delta k_1,\Delta k_2,\Delta k_3)$ plotted along the diagonal $\bar{g}^{(4)}_{BB}(\Delta k,\Delta k,\Delta k)$ for $n=0.31(12)$.  The dashed line shows a gaussian fit. The dotted lines bounding the shaded region show the fits to the statistical noise above and below each data point, as shown by the error bars. }
	\label{g4_diag}
\end{figure}

The lack of statistics in this dataset also prevented us from extracting the complete 4th and higher-order correlation functions.  However, by combining the 5 highest count rate data into a single dataset (with $n=0.31(12)$) we are able to reconstruct the maximally correlated case of two atoms on each side of the halo $\bar{g}_{BB}^{(4)}(\Delta k_1,\Delta k_2,\Delta k_3)$.  The four dimensional nature of this correlation function and the poor signal to noise make a clear visual representation problematic, so for simplicity we only plot the diagonal case of $\Delta k_1 = \Delta k_2 = \Delta k_3 = \Delta k$, shown in Fig.~\ref{g4_diag}.

\subsubsection*{Correlation function fits} 

To extract the maximum correlation amplitudes [$\bar{g}^{(2)}_{BB}(0), \bar{g}^{(3)}_{BB}(0,0)$ \textit{etc.}], the individual correlation functions are fitted with a 1D Gaussian fit function \cite{Hodgman2011,Dall2013_idealnbody}.  For $\bar{g}^{(2)}_{BB}(\Delta k)$, this is given by 
\begin{equation}\label{eq:g2bb_fit}
	\bar{g}^{(2)}_{BB}(\Delta k)=A e^{-\Delta k^2/2\sigma_{BB}^2}+O, 
\end{equation} 
where the fit parameters $A$, $\sigma_{BB}$ and $O$ correspond to the correlation amplitude, two-atom correlation length and offset respectively.  Note that this means that $\bar{g}^{(2)}_{BB}(0)= A+O$.  Similar fit functions are defined for the other correlation functions.  

For the three-atom correlation functions, we fit the 1D correlation functions to $\Delta k_1=\Delta k_2$ [see Fig.~
2 (g) and (h)], while the $\bar{g}^{(2)}_{BB}(0)$ values shown in Fig.~
3 (c) are taken from the average of fits $\bar{g}^{(3)}_{BB}(0,\Delta k_2\gg \sigma_{BB})$ and $\bar{g}^{(3)}_{BB}(\Delta k_1\gg \sigma_{BB},0)$ [see Fig.~
2\,(g) and (h) of the main text].  Due to poor statistics, the uncorrelated values $\bar{g}^{(3)}_{CL}(\Delta k_1\gg \sigma_{BB},\Delta k_2\gg \sigma_{BB})$ $\bar{g}^{(4)}_{BB}(\Delta k_1\gg \sigma_{BB},\Delta k_2\gg \sigma_{BB},\Delta k_3\gg \sigma_{BB})$ are also slightly larger than one.

Due to the angular integration inherent in our definition of the correlation functions (see, e.g., Eq.~(\ref{eq:g2cl})), smaller values of $\Delta k$ will have relatively fewer pairs in both the numerator and denominator of the correlation function compared to larger values of $\Delta k$ (the same is also true of $\Delta k_1$ and $\Delta k_2$).  This means the relative noise will be larger for small values of $\Delta k$.  To account for this, when fitting correlation functions to fits such as Eq.~(\ref{eq:g2bb_fit}), we perform three fits: one to all data, one excluding the smallest value of $\Delta k$ and one excluding the two smallest values of $\Delta k$.  In the data shown in Fig.~\ref{g2_fits}, the average of each fit parameter is then used, while the spread is taken as the fit uncertainty, which can be quite large for some data. 

\begin{figure}[bp]	\includegraphics[width=0.47\textwidth, keepaspectratio=true]{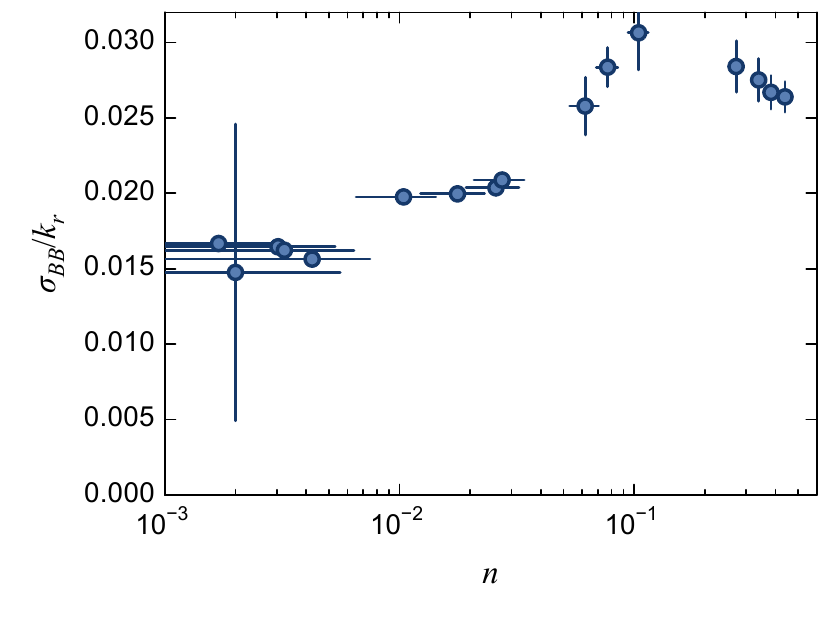}
	\caption{ {\bf $g^{(2)}_{BB}$ correlation length.}  Back-to-back correlation length $\sigma_{BB}$  extracted from Gaussian fits to the individual correlation plots, as shown in Fig.~\ref{g2_fits} (see text for details), plotted as a function of the average halo mode occupancy $n$, in units of the halo radius in momentum space ($k_{r}$).  Errorbars indicate combined statistical and fit errors.  }
	\label{g2_corr_widths}
\end{figure}

\subsubsection*{$g^{(2)}_{BB}(\Delta k)$ correlation length}

From the Gaussian fits shown in Fig.~\ref{g2_fits}, we can extract the two-particle back-to-back correlation length $\sigma_{BB}$ from Eq.~(\ref{eq:g2bb_fit}), which is plotted against average mode occupancy in Fig.~\ref{g2_corr_widths}.  The correlation length is seen to increase with higher mode occupancy.  Theoretically, we would expect \cite{Oegren2009} the length scale of $\sigma_{BB}$ to be set by the Thomas-Fermi radius $R_{\mathrm{TF}}$  of the source condensate that is being split into two equal halves in the Bragg diffraction regime. Even in the Kapitza-Dirac regime of splitting into multiple pairs of colliding condensates, the characteristic size of any particular pair is set by $R_{\mathrm{TF}}$ of the source condensate prior to any diffraction, and therefore $\sigma_{BB}$ should be approximately constant for halos of different average mode occupancy $n$. On the other hand, $\sigma_{BB}$ and $n$ have both previously been predicted (using numerical simulations \cite{Oegren2009}) to increase with increasing collision times, which is consistent with our observed result of increasing $\sigma_{BB}$ with larger $n$ (Fig.~6).  However, due to subtle differences in our experimental regime (e.g., in our setup $n$ is the final mode occupation, which is varied due to the different total number of atoms in a given pair of colliding condensates, rather than by changing the collision duration), a direct comparison is not possible.  A complete understanding of the relationship between $\sigma_{BB}$ and $n$ for our experiment would require detailed numerical simulations, which are beyond the scope of this work.

Note that the correlation lengths we measure are smaller than the experimental values for previously published work \cite{Perrin2007,Perrin2008}.  Small correlation lengths are desirable in this system, as the correlation length determines the halo mode volume, and thus smaller correlation lengths enable lower mode occupancies to be accessed.  The only caveat to this is that the correlation lengths should be larger than the detector resolution, as otherwise it will be impossible to differentiate between adjacent modes.  In our system we ensured that $\sigma_{BB}$ is always at least 3 times larger than our worst detector resolution.

\subsection*{2.4 General applicability of momentum microscopes to other many-body systems}

As mentioned in the main text, additional considerations may be necessary when looking to apply a similar momentum microscope to study other many-body systems.  In some strongly interacting systems the TOF expansion may not initially be ballistic, so in these cases alternative experimental steps or further minimal input from theory might be required, such as a treatment within short-time expansion or linear response theory, to relate the experimentally measured quantities to the in-trap momentum or quasi-momentum distributions. An example of the latter case is the expansion of a 1D Lieb-Liniger gas, for which the long-time asymptotic density distribution can be mapped to the in-trap distribution of quasi-momenta (also known as Bethe rapidities) \cite{Campbell:2015}.

\subsection*{2.5 Cauchy-Schwarz violation}

There are a number of subtleties which need to be considered when comparing our Cauchy-Schwarz violation to experiments with photons.  Firstly, in a quantum optics experiment the source is usually parametric down conversion in the low-gain regime, which produces a state approximating a twin-photon Fock state: $|\psi\rangle \sim |0,0\rangle + \alpha|1,1\rangle$, with $\alpha$ a constant.  Although there are higher occupancy states ($|22\rangle, |33\rangle$...) generated, in practice they occur with such low probability that any direct experimental measurement of $g^{(2)}_{CL}(0)$ would not result in any statistically significant signal.  This means that $g^{(2)}_{CL}(0)$, which enters into the definition of the correlation coefficient $C_2$ that quantifies the degree of Cauchy-Schwarz violation, can only accurately be inferred, for example from the visibility $V_{HOM}$ of a Hong-Ou-Mandel dip \cite{HOM,Lewis-Swan2014}, where 
\begin{equation}\label{eq:HOM_vis}
	V_{HOM}=1-\frac{1}{1+C_2}.  
\end{equation}
To the best of our knowledge, the largest $V_{HOM}$ reported is $V_{HOM}\approx 0.98$ ($98\%$ visibility) \cite{Lee2006}, which from Eq.~(\ref{eq:HOM_vis}) would imply a value of $C_2\approx 58$, hence the claim that our Cauchy-Schwarz violation is the best on record.  Note that for another experiment to beat our value of $C_2 >100$ it would need to measure $V_{HOM}>0.99$ ($>99\%$ visibility).  The high degree of violation that we observe clearly demonstrates the quantum nature of our halo.

All previous similar collision halo experiments with ultracold atoms were only able to measure peak correlation amplitudes $\bar{g}_{BB}^{(2)}(0) \!\simeq \!\bar{g}_{CL}^{(2)}(0)$ \cite{Perrin2007,Kheruntsyan2012}.  Therefore they were only able to demonstrate a violation of the Cauchy-Schwarz inequality using volume-integrated atom numbers, rather than bare peak correlations. This was made possible by the larger correlation volume of $g^{(2)}_{BB}$ in their system, which was due to the fact that the colliding sources were phase-fluctuating quasicondensates, rather than nearly pure condensates as is the case in our experiment.

\end{document}